\documentclass[sigplan,screen,balance=true]{acmart}

\usepackage{listings}
\usepackage{xspace}
\usepackage{xcolor}
\usepackage[switch]{lineno}
\usepackage{multirow}
\usepackage{bbding}
\usepackage[font=small]{subfig}
\usepackage{booktabs}
\usepackage{enumitem}

\newcommand{\SimName}{\textsc{Parendi}\xspace}

\newcommand{\shortparagraph}[1]{\indent\textbf{\textit{#1}.}\sloppy}

\newcommand{\figRef}[1]{\hyperref[fig:#1]{Fig.~\ref{fig:#1}}}
\newcommand{\secRef}[1]{\hyperref[sec:#1]{\S\ref{sec:#1}}}
\newcommand{\tabRef}[1]{\hyperref[tab:#1]{Table~\ref{tab:#1}}}
\newcommand{\codeRef}[1]{\hyperref[lst:#1]{Listing~\ref{lst:#1}}}
\newcommand{\equRef}[1]{\hyperref[lst:#1]{Equ.~\ref{eq:#1}}}

\newcommand{\textcode}[1]{\texttt{#1}}

\newcommand{\bmark}[1]{\textcode{#1}}
\newcommand{\smallrocket}[1]{\textcode{sr{#1}}}
\newcommand{\largerocket}[1]{\textcode{lr{#1}}}

\newcommand{\mIntel}{\textcode{ix3}\xspace}
\newcommand{\mAMD}{\textcode{ae4}\xspace}

\newcommand{\fiber}{fiber\xspace}
\newcommand{\fibers}{fibers\xspace}

\newcommand{\Fiber}{Fiber\xspace}
\newcommand{\Fibers}{Fibers\xspace}
\newcommand{\thread}{process\xspace}
\newcommand{\threads}{processes\xspace}

\definecolor{commentgreen}{RGB}{2,112,10}
\definecolor{eminence}{RGB}{108,48,130}
\definecolor{weborange}{RGB}{255,165,0}
\definecolor{frenchplum}{RGB}{129,20,83}
\definecolor{changecolor}{RGB}{255, 64, 0}
\newif\ifcomments

\commentsfalse

\ifcomments
\newcommand{\me}[1]{\noindent{\color{red} {\bf \fbox{Mahyar}} #1}}
\newcommand{\jl}[1]{\noindent{\color{orange} {\bf \fbox{Jim}} #1}}
\newcommand{\tb}[1]{\noindent{\color{blue} {\bf \fbox{Thomas}} #1}}
\newcommand{\thomas}[1]{\noindent{\color{blue} {\bf \fbox{Thomas}} #1}}
\newcommand{\sk}[1]{\noindent{\color{green} {\bf \fbox{Sahand}} #1}}
\else
\newcommand{\me}[1]{}
\newcommand{\jl}[1]{}
\newcommand{\tb}[1]{}
\newcommand{\thomas}[1]{}
\newcommand{\sk}[1]{}
\fi

\makeatletter
\DeclareRobustCommand{\change}{%
  \@bsphack
  \color{changecolor}%
  \@esphack
}
\DeclareRobustCommand{\stopchange}{%
  \@bsphack
  \normalcolor
  \@esphack
}
\makeatother

\definecolor{commentgreen}{RGB}{2,112,10}
\definecolor{eminence}{RGB}{108,48,130}
\definecolor{weborange}{RGB}{255,165,0}
\definecolor{frenchplum}{RGB}{129,20,83}
\definecolor{ForestGreen}{RGB}{34,139,34}
\definecolor{BrickRed}{RGB}{170,74,68}
\definecolor{RoyalBlue}{RGB}{65, 105, 225}

\copyrightyear{2025}
\acmYear{2025}
\setcopyright{cc}
\setcctype{by}
\acmConference[ASPLOS '25]{Proceedings of the 30th ACM International Conference on Architectural Support for Programming Languages and Operating Systems, Volume 2}{March 30-April 3, 2025}{Rotterdam, Netherlands}
\acmBooktitle{Proceedings of the 30th ACM International Conference on Architectural Support for Programming Languages and Operating Systems, Volume 2 (ASPLOS '25), March 30-April 3, 2025, Rotterdam, Netherlands}\acmDOI{10.1145/3676641.3716010}
\acmISBN{979-8-4007-1079-7/25/03}

\begin{document}

\title{\SimName: Thousand-Way Parallel RTL Simulation}

\author{Mahyar Emami}
\email{mahyar.emami@epfl.ch}
\affiliation{%
    \institution{EPFL}
    \city{Lausanne}
    \country{Switzerland}
}

\author{Thomas Bourgeat}
\email{thomas.bourgeat@epfl.ch}
\affiliation{%
    \institution{EPFL}
    \city{Lausanne}
    \country{Switzerland}
}

\author{James R. Larus}
\email{james.larus@epfl.ch}
\affiliation{%
    \institution{EPFL}
    \city{Lausanne}
    \country{Switzerland}
}



\begin{abstract}
Hardware development critically depends on cycle-accurate RTL simulation.
However, as chip complexity increases, conventional single-threaded simulation becomes impractical due to stagnant single-core performance.

\SimName is an RTL simulator that addresses this challenge by exploiting the abundant fine-grained parallelism inherent in RTL simulation and efficiently mapping it onto the massively parallel Graphcore IPU (Intelligence Processing Unit) architecture.
\SimName scales up to 5888 cores on 4 Graphcore IPU sockets. It allows us to run large RTL designs up to 4$\times$ faster than the most powerful state-of-the-art x64 multicore systems.

To achieve this performance, we developed new partitioning and compilation techniques and carefully quantified the synchronization, communication, and computation costs of parallel RTL simulation:
The paper comprehensively analyzes these factors and details the strategies that \SimName uses to optimize them.
\end{abstract}

\maketitle

\section{Introduction}

Hardware developers spend as much as a quarter of their time \emph{running} simulations~\cite{SiemensVerifStudyPart8Asic,SiemensVerifStudyPart4Fpga}.
Cycle-accurate RTL (Register Transfer Level) simulation is an essential tool for debugging and validating an ASIC or FPGA design, but it can be time-consuming to run.

\begin{figure}[h!]
    \centering
    \includegraphics[width=0.95\columnwidth]{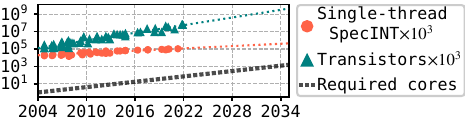}
    \caption{Chip growth and single-thread performance~\cite{processor_scaling_data}.
 The dashed line predicts the core count, assuming linear scaling, necessary to simulate a state-of-the-art chip at the same rate as in 2006.
 }
    \label{fig:sim_rate_projection}

\end{figure}

Unfortunately, its slow speed hampers the design process.
\figRef{sim_rate_projection} shows the increasing gap between single-thread performance and package transistor count.
It shows that the single-thread simulation of the new generations of chips on existing computers is becoming less feasible.

One appealing solution is to exploit the inherent parallelism of RTL designs by simulating them on parallel computers~\cite{manticore_arxiv,repcut,rtlflow,DBLP:conf/date/LopezParadisLAWMB23}.
However, \figRef{sim_rate_projection} shows that simulating today's chips at the same rate as we simulated chips in 2006 requires parallel simulation that can utilize hundreds or thousands of cores.

This paper presents a practical solution to the problem of parallelizing RTL simulation of large (e.g., 100-core SoCs) across a few \emph{thousand} cores.
To demonstrate, we build an RTL simulator running on the Graphcore IPU~\cite{graphcore,DBLP:journals/corr/abs-1912-03413}, a 1472-core chip that is the building block of parallel machine-learning systems.
Although the IPU is not well known, its architecture contains many features--high core count, fast synchronization, and high internal and external bandwidth--that are especially well-suited for large-scale RTL simulation.

A parallel RTL simulator on a massively parallel machine must balance synchronization, communication, and computation.
We analyze these factors to clarify their relations.
These axes are not independent, which makes it challenging to partition an RTL simulation across many cores.

We use these experimental insights to build %
\SimName\footnote{\SimName is the female Zoroastrian angel of abundance.},
the first scalable, multi-thousand-way parallel RTL simulator.
\SimName is open-source, facilitating further research.
For large designs, \SimName demonstrates performance and efficiency gains across multiple dimensions.
It runs up to 4$\times$ faster than multithreaded Verilator (the fastest RTL simulator).
In nightly cloud testing scenarios, \SimName could reduce costs by more than 2$\times$.
It also compiles large designs 12$\times$ faster and uses 18$\times$ less memory than Verilator.

The contributions of this work are:
\begin{itemize}[leftmargin=*]
    \item A quantitative study of massively parallel RTL simulation.
    \item A new communication- and duplication-aware compilation strategy for large-scale RTL simulation.
    \item Implementation of \SimName, the first open-source\footnote{
        \href[]{https://github.com/epfl-vlsc/parendi}{https://github.com/epfl-vlsc/parendi}
 } RTL simulator that runs on thousands of cores.
    \item An evaluation of \SimName on a Graphcore system with 5888 cores that shows that it cost-effectively exceeds Verilator's performance on a high-end x64 system by up $4.0\times$.
\end{itemize}

The paper is organized as follows:
\secRef{background} provides background on the IPU.
\secRef{parallel_sim} details the parallel RTL simulation strategy.
\secRef{bsp_perf_discussion} presents a high-level performance study.
\secRef{compiler} outlines \SimName's design.
\secRef{evaluation} evaluates \SimName on IPU and Verilator on x64.
\secRef{related} reviews related work.
\secRef{conclusion} concludes.

\begin{figure}
    \centering
    \includegraphics[width=\columnwidth]{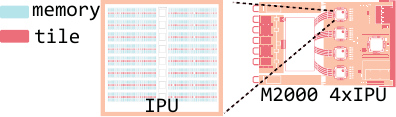}
    \caption{The IPU processor and M2000 server blade.}
    \label{fig:graphcore_ipu}
\end{figure}
\section{Graphcore IPU}\label{sec:background}

A Graphcore IPU is a single package containing 1472 tiles (physical cores) connected by a high-bandwidth network (IPU exchange) with 11~TiB/s all-to-all bandwidth~\cite{graphcore}.
The IPU is a multiple-instruction, multi-data (MIMD) architecture in which each tile runs an independent instruction stream. By contrast, GPUs use SIMD or SIMT execution, where groups of threads (Warps) simultaneously execute the same instruction on different data.
The IPU is a message-passing machine.
Each tile can only access its private memory and must explicitly communicate through the exchange fabric.
An IPU has a total on-chip memory of approximately 900~MiB, with each tile having exclusive access to 624~KiB.

\figRef{graphcore_ipu} displays a Graphcore M2000 IPU server, a 1U unit housing 4 IPUs with a 320~GiB/s exchange fabric, totaling 5888 tiles.
Systems can scale to 16 or 64 IPUs using multiple boards. Our study utilized a single board.

An IPU system (with any number of IPUs) is programmed in C++ using the bulk synchronous parallel (\textbf{BSP})~\cite{valiant_bridging_1990} programming model, supported by the \emph{poplar} SDK~\cite{poplar_sdk} and clang-based C++ compiler \emph{popc}.
The IPU directly supports BSP communication and synchronization.
The following section describes BSP and how to apply it to RTL simulation.

\section{Parallel RTL Simulation}\label{sec:parallel_sim}
Hardware description languages (HDL) like Verilog express digital sequential circuits.
An HDL program contains stateful, clocked elements called registers interconnected by wires and stateless combinational logic.
Register transfer level (RTL) is a set of clocked registers and combinational logic.

This paper considers cycle-accurate RTL simulation, where combinational logic has zero delay.
Also, we only use full-cycle (activity-oblivious) simulation, which evaluates an entire circuit at each RTL cycle.
The alternative is event-driven (activity-aware) simulation.
In general, full-cycle simulators perform better---sometimes by orders of magnitude---because tracking value changes in RTL is expensive~\cite{DBLP:conf/dac/BeamerD20}.

\subsection{Shared-Memory Simulation}

Parallel RTL simulation poses challenges for cache-coherent, shared-memory computers.
First, fine-grained parallelism requires frequent synchronization, which is costly on a shared-memory multiprocessor~\cite{manticore_arxiv,repcut}.
Second, the RTL tasks perform fine-grained, point-to-point communication.
In an RTL design, a task may communicate only a few bytes of data to the known cores computing its neighbors in the RTL graph, but all transfers go through the last-level cache (LLC).
Finally, when compiled into code, RTL can have a high data and instruction reuse distance, which makes caches perform poorly.
Most data items and instructions are accessed once per simulated RTL cycle, which might span millions of machine cycles.
When large designs do not fit the caches, these memory references incur cache misses each RTL cycle~\cite{DBLP:conf/micro/Zhou0LWH23}.

\subsection{BSP RTL Simulation}\label{sec:bsp_sim}

To alleviate the first two problems, we use Valiant's bulk synchronous parallel (\textbf{BSP})~\cite{valiant_bridging_1990} model.
BSP is a message-passing model alternating two phases: (i) computation and (ii) communication.
In the computation phase, parallel \emph{processes} run, reading shared values and modifying only private data.
Computation ends at a barrier.
Then, the communication phase transfers newly computed private values to consuming processes.
Communication also ends at a barrier, after which the next computation phase begins.
The appeal of BSP is that it reduces synchronization to two global barriers per RTL clock cycle.
\footnote{
Other parallel RTL simulation systems utilize this computation model~\cite{DBLP:conf/aspdac/NanjundappaPJS10,DBLP:conf/dac/VincoCBF12,DBLP:conf/dac/ChatterjeeDB09, DBLP:conf/date/ChatterjeeDB09,DBLP:journals/todaes/ChatterjeeDB11,repcut,manticore_arxiv}, but it was only recently called out as \emph{BSP}~\cite{manticore_arxiv}.
}

\begin{figure}
    \includegraphics[width=\columnwidth]{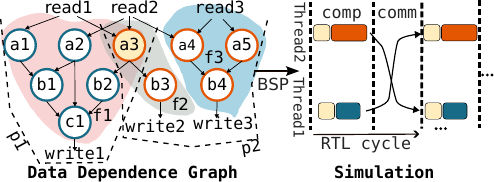}
    \caption{BSP Simulation of an RTL data dependence graph. The graph contains three fibers (\textcode{f1}, \textcode{f2}, \textcode{f3}), partitioned into two processes (\textcode{p1}, \textcode{p2}), running on two threads. \textcode{a3} is duplicated. The run on the right shows the computation and communication phases, separated by barriers.}
    \label{fig:bsp_sim}

\end{figure}

\figRef{bsp_sim} contains a sample \emph{data dependence graph} RTL circuit.
This graph splits each RTL register into two values: a read-only value (\emph{current}, at the leading clock edge) and a write-only value (\emph{next}, at the end).
The \emph{current} values at the top (e.g., \textcode{read1}) are fed into stateless combinational logic (circles) that computes the \emph{next} register values (e.g., \textcode{write1}).

The dashed lines in~\figRef{bsp_sim} \emph{partition} the graph into BSP \threads \textcode{p1} and \textcode{p2}.
Each \thread reads a set of RTL registers and computes new values for one or more (e.g., \textcode{write1} in \textcode{p2}).
Since we only communicate register values at the end of the computation phase, \threads may need to duplicate intermediate computations (e.g., \textcode{a3} is in both \textcode{p1} and \textcode{p2}).

The right side of~\figRef{bsp_sim} shows the evaluation of the example.
The parallel \threads synchronize at a barrier (dashed vertical lines).
The first barrier marks the end of the computation, after which we exchange \emph{next} values (\textcode{write1}, \textcode{write2}, and \textcode{write3}) and update the \emph{current} values accordingly (\textcode{read1}, \textcode{read2}, and \textcode{read3}).
We finish the communication with a barrier and conclude one RTL simulation cycle.

A parallel RTL compiler partitions the data dependence graph to minimize the time spent running a simulation cycle.
\textcode{p1} and \textcode{p2} are not the only possible partitions of~\figRef{bsp_sim}.
We call the atoms of BSP simulation \emph{\fibers}.
A \fiber is the smallest set of operations that uniquely produces the \emph{next} value of a single register.
\figRef{bsp_sim} contains three registers and three \fibers \textcode{f1} to \textcode{f3}, partitioned into \textcode{p1} = \{\textcode{f1}\} and \textcode{p2} = \{\textcode{f2}, \textcode{f3}\}.

It is worth noting that BSP is only one of many possible parallel simulation techniques.
In BSP, nodes \textcode{a1} and \textcode{a2} belong to the same \fiber, so they run one after the other.
We could consider a fine-grain parallel execution in which individual nodes in~\figRef{bsp_sim} evaluate in parallel, with point-to-point synchronization.
Verilator~\cite{snyder_mt_2018,snyder_mt_2019,snyder_verilator_2020} uses this approach~\cite{Sarkar86}.
The advantage of fine-grained parallelism is avoiding duplicated work at the cost of more synchronization.

\section{Analysis of BSP RTL Simulation}\label{sec:bsp_perf_discussion}

We now measure and analyze the principal performance factors in BSP RTL simulation using small benchmarks on the M2000 quad-IPU system and an Intel Xeon Gold 6348 56-core dual-socket processor (see~\tabRef{hw_setup} for details).

Parallel performance depends on synchronization, communication, and computation costs.
In BSP, the sum of the three is the time to simulate one cycle, so the simulation rate (in a thousand RTL cycles per second or kHz) is

\begin{equation}\label{eq:sim_rate}
    r_{cycle} = \frac{1}{t_{sync} + t_{comm} + t_{comp}},
\end{equation}

\noindent where $t_{sync}$, $t_{comm}$, and $t_{comp}$ are per RTL cycle synchronization, communication, and computation latencies.
Reducing this sum increases the simulation rate. Below, we explore how each term behaves as we seek to increase parallelism.
Our analysis reveals salient architectural features of the IPU and x64, providing insight into compilation strategies.

\subsection{Synchronization}\label{sec:bsp_t_sync}

BSP requires two global synchronizations per simulated clock cycle, so $t_{sync}$ is the cost of two barriers.
Therefore, $t_{sync}$ is independent of the simulated design.
However, since the cost of a barrier increases with parallelism, it depends on the number of hardware threads used for a simulation.

To explore the relationship between $t_{sync}$ and performance, we simulate a set of pseudo-random number generators (PRNG), each performing three XORs and three shifts~\cite{xorshift}.
The simulated PRNGs are independent; so, $t_{comm} = 0$, but $t_{sync} > 0$ as we still need to synchronize with the RTL clock.
Therefore, if $t_{sync}$ is small compared to the computation cost, we expect to observe a near-constant simulation rate.

We simultaneously increase the number of PRNGs and computation units—tiles (IPU) or threads (x64).
In each set of experiments, we keep the amount of work per tile (thread) constant.
Note that each PRNG consists of one \fiber, but we can sequentially execute multiple \fibers on one tile (thread) to vary the computation-to-synchronization cost ratio.

We use the IPU's built-in barrier and sweep the tiles from 64 to 5888 by 64.
On x64, we use a user-space (atomic fetch-and-add) barrier, measuring from 1 to 56 threads.
This type of barrier performed better than OpenMP's built-in, MCS, or sense barriers.

\begin{figure}
    \centering
    \includegraphics[width=\columnwidth]{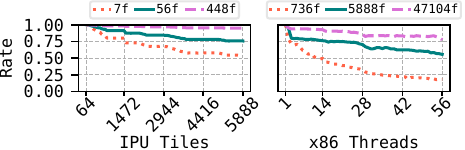}
    \caption{IPU and x64 PRNG rates}
    \label{fig:rate_prngs}

\end{figure}

\figRef{rate_prngs} shows the measured rate on the IPU (normalized to the rate with 64 tiles) and on x64 (normalized to the rate with one thread).
Each line shows the performance with a fixed quantity of \fibers per tile: 7, 56, and 448 on the IPU and 736, 5888, and 47104 on the x64.
The total work with 5888 tiles on IPU is the same as 56 threads on x64.
We normalized each experiment to itself as we are not comparing absolute simulation rates between the machines.

With 7 \fibers per tile on the IPU, synchronization causes performance to fall by almost $50\%$ as the number of tiles increases.
Synchronization latency becomes less detrimental as the computation per tile increases (with 448 \fibers, performance falls by a few percent).
The cost of synchronization on x64 is high, even with many \fibers per thread.
With 736 \fibers per thread, performance drops by more than $75\%$, and even with 47104 \fibers per thread, performance falls by 25\%.
The IPU has a native hardware barrier that consumes only a few hundred IPU cycles.
By contrast, x64 barrier synchronization requires expensive atomic memory accesses that could require a few thousand cycles with 56 threads.

\figRef{rate_prngs} reveals a simple rule-of-thumb.
Masking synchronization overhead on x64 requires hundreds of thousands of instructions per thread (each \fiber is roughly 6 instructions), whereas on the IPU, a few thousand instructions adequately hide synchronization overhead.
The IPU supports very fine-grain parallelism, whereas the x64 does not.

\subsection{Communication}\label{sec:bsp_t_comm}

Similar to $t_{sync}$, we expect $t_{comm}$ to increase as we increase parallelism, as adding tiles (threads) means more values are communicated among tile (thread).
Unlike synchronization, communication depends on the specifics of an RTL design and the partition of \fibers among tiles (threads).
We summarize these considerations into two parameters: bytes sent from each tile ($b$) and number of tiles in the simulation ($m$).

To first order, we might expect $t_{comm} = \frac{m\times b}{bw}$, where $bw$ is the communication bandwidth and $m \times b$ is the total communication volume.
Additional parallelism can increase communication latency if it increases the inter-tile (thread) volume.
Therefore, at some point, the increase in $t_{comm}$ could outweigh the benefits of spreading computation among more tiles (threads).
Alternatively, $t_{comm}$ could be almost independent of $m$ and depends primarily on $b$.
In this case, performance would increase \emph{monotonically} with parallelism.

We found that communication within a single IPU appears to depend primarily on $b$, but communication between IPUs depends on $m \times b$.
We demonstrate this with two experiments.

\begin{figure}[h]
    \centering
    \includegraphics[width=0.85\columnwidth]{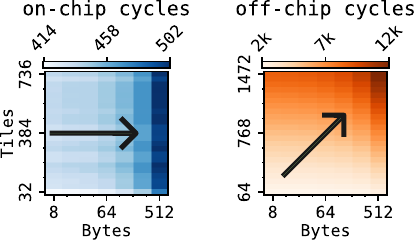}
    \caption{Measured communication cycles on the IPU}
    \label{fig:ipu_bw}
\end{figure}

First, consider $2m$ tiles running on one IPU.
We \emph{randomly} partition the $2m$ tiles into two sets of $m$ tiles and send a fixed number of bytes in both directions between the sets.
The left plot in~\figRef{ipu_bw} shows the measured IPU cycle counts (averaged over 10 random bi-partitions).
The cycle counts include $t_{sync}$ as an exchange requires synchronization.
The horizontal axis is the number of bytes each tile sends and receives ($b$).
The vertical axis is the number of tiles ($m$).
The on-chip $t_{comm}$ increases only in the direction of increasing $b$ as shown by the arrow in the left chart of~\figRef{ipu_bw}.

In the second experiment, one tile in each pair resides on one IPU and the other on another IPU, so all traffic goes off-chip.
The right chart in~\figRef{ipu_bw} reports the results.
It shows a vastly different behavior: $ t_{comm}$ increases with increasing parallelism \emph{and} bytes per tile as it depends on $m\times b$ (the diagonal arrow delineates the direction of change).
Furthermore, the increase is more pronounced.

At the plots' darkest points, we consume $13\%$ and $82\%$ of the maximum measured communication on- and off-chip bandwidth, respectively (7.7~TiB/s and 107~GiB/s).
The on-chip experiment is far from saturating the bandwidth, so latency is insensitive to tile count.
By contrast, the off-chip experiment runs near the fabrics's maximum bandwidth, so additional communication increases contention and latency.

In conclusion, the difference between these communication fabrics means that minimizing off-chip communication volume is a first-class concern when the traffic is large.

It is worth noting a limitation of Graphcore BSP communication.
The IPU's exchange fabric is \emph{statically scheduled}; hence, communication \emph{must} start with a barrier to ensure all tiles are at the same point in execution.
Unfortunately, this precludes optimization such as overlapping computation and communication or dynamic load balancing.

\subsection{Computation}\label{sec:bsp_t_comp}

At first glance, optimizing $t_{comp}$ is similar to the \emph{multiprocessor independent task scheduling} (\emph{makespan minimization}) problem~\cite{DBLP:journals/ior/GareyGJ78}.
In this classic problem, we consider a set of tasks (\fibers) $F = \lbrace f_1, ..., f_n \rbrace$ with corresponding execution times $t_i, ..., t_n$.
The goal is to schedule these tasks on $m$ tiles (threads) to minimize the longest execution time across all tiles.
This problem is NP-hard~\cite{mpsched_np}, but polynomial-time approximations exist~\cite{DBLP:journals/jacm/Sahni76,DBLP:journals/ior/GareyGJ78}.

In the classical problem, tasks (\fibers) have fixed execution times, independent of where they run.
However, in RTL simulation, two \fibers might compute a shared intermediate value (for example, value \textcode{a3} in \figRef{bsp_sim}).
Collocating these two fibers in the same process enables optimization.
However, it complicates the partitioning problem.
Each \fiber consists of a set of computation nodes (the nodes in~\figRef{bsp_sim}).
If we denote the execution time of a BSP \thread using $\tau(.)$, then a \thread made up of \fibers $f_i$ and $f_j$ would have $\tau(f_i \cup f_j) = t_i + t_j - \tau(f_i \cap f_j)$ since we need to execute the shared code only once.
Moreover, merging \fibers eliminates communication ($t_{comm}$) so $\tau(f_i \cup f_j) \leq t_i + t_j - \tau(f_i \cap f_j)$.
This is a submodular function, and this variant of the scheduling problem is called submodular load balancing (SLB).
SLB is inherently more complex than classic scheduling and challenging to get even modest approximation guarantees ($\sqrt{{n} / {log(n)}}$~\cite{SLB}).

\begin{figure}
    \centering
    \subfloat[Stragglers impose a lower bound on $t_{comp}$. \Fibers sorted for visualization.]  {
        \label{fig:ipu_stragglers}
        \includegraphics[width=0.8\columnwidth]{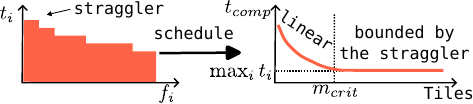}
    } \\
    \subfloat[\Fiber computation cycles in \bmark{pico}, \bmark{bitcon}, and \bmark{rocket}.]  {
        \label{fig:ipu_fiber_plot}
        \includegraphics[width=\columnwidth]{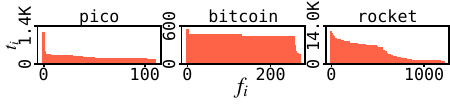}
    } \\
    \subfloat[Reducing $t_{comp}$ through parallel execution (base-2 log scale). Dashed lines show a perfect scaling.]  {
        \label{fig:ipu_compute_cycles_plot}
        \includegraphics[width=0.9\columnwidth]{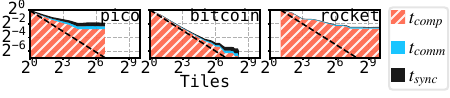}
    }
    \caption{Straggler \fibers and performance scaling regions.}
\end{figure}

\begin{table}
    \centering
    \resizebox{1\columnwidth}{!}{%
        \setlength{\tabcolsep}{2pt}
        \begin{tabular}{llllllllllll}
            \toprule
            \multicolumn{6}{c}{\bf \SimName}     & \multicolumn{6}{c}{\bf Verilator on \mIntel}                                                                                                                                                                                                                                                                                                       \\
            \cmidrule(l{4pt}r{4pt}){1-6}
            \cmidrule(l{4pt}r{4pt}){7-12}
            \multicolumn{2}{c}{\bf \bmark{pico}} & \multicolumn{2}{c}{\bf \bmark{bitcoin}}      & \multicolumn{2}{c}{\bf \bmark{rocket}} & \multicolumn{2}{c}{\bf \bmark{pico}} & \multicolumn{2}{c}{\bf \bmark{bitcoin}} & \multicolumn{2}{c}{\bf \bmark{rocket}}                                                                                                                                    \\

            {\bf par.}                           & {\bf kHz}                                    & {\bf par.}                             & {\bf kHz}                            & {\bf par.}                              & {\bf kHz}                              & {\bf par.} & {\bf kHz}                    & {\bf par.} & {\bf kHz}                    & {\bf par.} & {\bf kHz}                   \\
            \cmidrule(l{4pt}r{4pt}){1-2}
            \cmidrule(l{4pt}r{4pt}){3-4}
            \cmidrule(l{4pt}r{4pt}){5-6}
            \cmidrule(l{4pt}r{4pt}){7-8}
            \cmidrule(l{4pt}r{4pt}){9-10}
            \cmidrule(l{4pt}r{4pt}){11-12}
            1                                    & 168.7                                       & 1                                      & 14.5                                & 2                                       & 17.7                                  & 1          & 14141.7                      & 1          & 537.4                       & 1          & 220.3                    \\
            111                                  & \textcolor{ForestGreen}{629.4}              & 270                                    & \textcolor{ForestGreen}{935.2}      & 1211                                    & \textcolor{ForestGreen}{93.3}         & 2          & \textcolor{BrickRed}{490.4} & 2          & \textcolor{BrickRed}{232} & 2          & \textcolor{BrickRed}{99.2} \\
            \bottomrule
        \end{tabular}
    }
    \caption{Simulation rate in kHz.
        \textbf{par} is the tile- or thread-count used to achieve the rate in \textbf{kHz}.
        See~\tabRef{hw_setup} for the technical specification of \mIntel.
        }
    \label{tab:bitcoin_picorv32_perf}

\end{table}

In the trivial case, when fewer tasks exist than tiles (threads) ($n \leq m$), the optimal solution is to assign a \fiber to each tile (thread).
It is impossible to improve $t_{comp}$ beyond $\max_i t_i$ as the slowest \fiber (the straggler) bounds $t_{comp}$ from below.
Encountering this bound on x64 hardware is unlikely: even relatively small designs have a few hundred \fibers, an order of magnitude more than available cores.
However, a single IPU chip has 1472 tiles, sufficient for small to medium-sized designs so that a straggler can limit IPU performance.

\figRef{ipu_stragglers} depicts the SLB problem: mapping \fibers to tiles results in a linear region in which $t_{comp}$ falls almost linearly with additional tiles.
The benefits become less significant, and we eventually plateau at $\max_i t_i$ (with sufficient tiles) with
$m_{crit}$ as the minimum tiles needed to this point.
To maximize the simulation rate, we only need $m_{crit}$ tiles; having more would not help as the straggler is the limit.

To put this into perspective, consider three small RTL designs:
(1) a multi-cycle RISC-V \bmark{pico} core~\cite{pico}, (2) a \bmark{bitcoin} miner~\cite{bitcoin}, and (3) a \emph{small} \bmark{rocket} pipelined RISC-V core~\cite{rocket_chip}.
These small designs contain more \fibers than x64 systems cores: 111, 270, and 1211 \fibers, respectively.
We run each benchmark using \SimName, described later in~\secRef{compiler}.

\figRef{ipu_fiber_plot} shows \fiber computation latency ($t_i$ for each $f_i$) of the three benchmarks (in IPU machine cycles).
\figRef{ipu_compute_cycles_plot} illustrates the corresponding scheduled execution times with the dashed diagonal representing a perfect linear reduction.
We normalize machine cycle counts to the minimum parallel execution: 1 tile in \bmark{pico} and \bmark{bitcoin}, 2 tiles in \bmark{rocket} (a single tile cannot hold sufficient code and state for \bmark{rocket}).

First, $t_{comp}$ in~\figRef{ipu_compute_cycles_plot} follows the trend of~\figRef{ipu_stragglers}: imbalanced \fibers yield a small linear scaling region.
\bmark{pico} is the most imbalanced and settles to a final $t_{comp}$ extremely quickly.
\bmark{rocket} is slightly more scalable but \bmark{bitcoin} performs the best as its \fibers are roughly balanced.
Second, $t_{sync}$ and $t_{comm}$ increase with additional tiles.
However, the $t_{comp}$ reduction is always larger, so the execution cost decreases.

\tabRef{bitcoin_picorv32_perf} compares the wall-clock simulation rate of the IPU, using \SimName, against an Intel Xeon 6348 (see~\tabRef{hw_setup} for details), using Verilator.
We show the simulation rate using a single tile (except for \bmark{rocket}, which needs more memory than available in one tile) and the maximum number of tiles, where we assign one \fiber per tile.
For x64, we report the single-thread and best multi-thread performance.

None of the three small benchmarks show any speedup on x64 from parallelism since the synchronization cost is too high (see~\secRef{bsp_t_sync}).
These results do not mean that Verilator cannot speed up RTL simulation.
In~\secRef {evaluation}, we show that Verilator does an excellent job of parallelizing code.
However, our analysis of $t_{sync}$ shows that a simulated design on the x64 must be large enough to mask synchronization overhead, and these three benchmarks need to be bigger.

Verilator's inability to scale these three benchmarks supports our claim that the straggler \fiber is not a performance limit on x64 as synchronization latency dominates.
On the other hand, stragglers are a fundamental concern for \SimName for small designs.
\tabRef{bitcoin_picorv32_perf} shows that \SimName's \emph{parallel} performance does not manage to even match Verilator's \emph{single-thread} performance for \bmark{pico} and \bmark{rocket}, despite modest gains from parallelism.
\SimName runs \bmark{bitcoin} using a single-tile at 14.5~kHz, far from Verilator's single-thread performance (537~kHz).
However, with 270 tiles, \SimName runs \bmark{bitcoin} at 935.2~kHz, faster than Verilator's single- and multi-thread performance.
Lastly, note that single-tile execution of \bmark{pico} and \bmark{bitcoin} on the IPU are approximately $84\times$ and $37\times$ slower than x64.
Consequently, the IPU has to significantly scale RTL simulation to reach Verilator's single-thread performance.
\section{\SimName Compiler}\label{sec:compiler}

\SimName is a Verilog compiler for the IPU systems.
It is derived from Verilator to take advantage of its optimizations and maturity.
However, \SimName contains significant changes that target the IPU (message-passing) rather than the x64 (shared memory).
\SimName also includes new scheduling and partitioning passes and IPU-specific optimizations.

\SimName generates a C++ BSP program that uses Graphcore's \emph{poplar} programming framework.
The code defines each tile's computation and how the tiles communicate.

At a high level, \SimName's primary responsibility is to partition RTL across the tiles of an IPU system.
The user specifies the number of tiles.
\SimName tries to maximize the simulation rate by finding an appropriate partitioning of \fibers to tiles.
We briefly describe our partitioning strategy.

\subsection{Partitioning}\label{sec:partitioning}

After generating a data dependence graph (see~\figRef{bsp_sim}), we find the \fibers by collecting the nodes that transitively feed into each sink node by crawling the graph in reverse.

Once we form the set of \fibers, we must solve the SLB problem (\secRef{bsp_t_comp}).
It is also crucial to recognize the interdependence between computation ($t_{comp}$) and communication latency ($t_{comm}$).
In addition, each tile has a finite memory.
Consequently, our partitioning algorithm must consider duplication, communication, and memory limitations.

We solve this problem in multiple steps, each pursuing a different goal.
At each algorithm step, we \emph{merge} \fibers into \emph{\threads}.
On the IPU, a \thread is a collection of \fibers that will eventually run on a tile (see \figRef{bsp_sim}).
There are four stages in our algorithm: (1) Reduce data memory footprint, (2) minimize off-chip communication, (3) reduce $t_{comm}$ while keeping $t_{comp}$ unchanged, (4) match the number of \fibers to the available hardware.

In the first stage, we merge \fibers that reference the same RTL array but only for \emph{very} large arrays (e.g., $\geq$ 128~KiB, tunable).
We do this to save memory at the cost of possibly increasing $t_{comp}$, so it is only worthwhile for large arrays.\footnote{
    Doing so reduces the probability of running out of tile memory later.
    Consider a design with one 256~KiB array and four 64-KiB ones.
    Assume each array references 3 equal-size {\fibers}:
    $a_1, a_2, a_3$ by the 256 KiB array, $b_1, b_2, b_3$ by the first 64-KiB array, ..., and $e_1, e_2, e_3$ by the fourth 64 KiB array.
    The goal is to create 3 balanced \threads.
    Because there are more 64-KiB arrays, we could end up merging fibers that reference distinct arrays, e.g., $\lbrace a_1, b_1, c_1, d_1, e_1\rbrace$.
    Such a \thread needs 512 KiB of data memory, which exceeds the available on-tile data memory.
    However, if we pre-merge $\lbrace a_1, a_2, a_3\rbrace$, in the worst case we will end up with a \thread such as $\lbrace a_1, a_2, a_3, b_1, c_1\rbrace$ requiring 384~KiB.
    No other balanced \thread would exceed 256~KiB either.
}

The second stage minimizes off-chip exchanges if \SimName is compiling for multiple IPUs.
We do this by partitioning a hypergraph of \fibers, in which hypernodes represent \fibers, and hyperedges represent RTL registers.
If two \fibers access the same register (read or write), their corresponding hypernodes share a hyperedge.
The hyperedge weights are the number of words in an RTL register, and hypernodes are unweighted.
For $k$ IPUs, we use the KaHyPar~\cite{DBLP:journals/jea/SchlagHGASS22} library to find a $k$-way balanced partition of the hypergraph that minimizes the \emph{cut} (hyperedges crossing the partitions).
KaHyPar produces $k$ roughly equally-sized sets of \fibers.
However, each set may contain more than 1472 \fibers, so we must further merge \fibers to fit the available tiles.

In the third stage, we \emph{conservatively} merge the smallest \fibers to reduce communication within each target IPU without increasing computation latency.
Intuitively, we move right to the left in the right subplot of~\figRef{ipu_stragglers} towards $m_{crit}$.
If we reduce the number of \fibers to fit the tiles in an IPU without crossing $m_{crit}$, then we can use the optimal $t_{comp} = \max_i t_i$ and a pseudo-optimal $t_{comm}$.
We create one \thread per \fiber and estimate its execution cost.
Recall from~\secRef{bsp_t_comp} that the cost of a \thread is submodular with respect to its \fibers.
We use a dense bitset data structure to represent duplication across \fibers and efficiently compute intersection and union in the submodular cost function---$\tau(f_i \cup f_j) = t_i + t_j - \tau(f_i \cap f_j)$.
Moreover, we use another bitset to track the memory usage after merging, accounting for deduplication.

In each iteration, we select the \thread with the shortest execution time and try to merge it with another with which it communicates, so long as their merged time does not surpass the worst existing execution time.
If we cannot perform the merge because of overflowing memory or exceeding the straggler execution time, we consider merging the two smallest \threads.
If that fails, we skip the candidate \thread and move to the next one.
We merge \threads until we process all of them or reach the desired tile count.

The final stage only runs if the third stage fails to reach the desired tile count.
We follow the same strategy as in the third stage but allow worst-case execution time to increase.
At the end of this stage, the number of \threads must fit the available hardware.
Otherwise, the compilation fails because the design is too large to fit the hardware resources.

\subsection{IPU-Specific Optimizations}
\SimName extends Verilator's optimizations with a few IPU-specific ones.
We briefly describe the most important ones.

\shortparagraph{Differential exchange}
RTL arrays are common in hardware designs, e.g., a register file or a cache bank.
If a \thread reads an array, it needs a full copy on its tile.
We avoid sending whole arrays by using static analysis to determine the number of updates to an array, though not their location or condition (e.g., a 2-port SRAM with byte-strobes).
With this analysis, we only send the changes instead of an entire array.

\shortparagraph{Aggressive block splitting}
We extend Verilator's \textcode{V3Split} pass, favoring parallelism over code bloat, to maximally split all clocked code blocks.

\shortparagraph{Aggressive inline}
\SimName ensures that the simulation program on the IPU is free of function calls.
Inlining can increase code size and produce excessive instruction cache pressure on x64, especially in RTL simulation, where nearly every instruction executes only once per RTL cycle, except for functions invoked multiple times.
An IPU tile has no instruction cache but a 624~KiB local memory, of which 200~KiB holds executable code.
So, a single IPU chip has $\approx$300~MiB of on-chip instruction memory space, which allows \SimName to aggressively inline code.

\subsection{Limitations}

Currently, \SimName only supports a subset of Verilator's clocking capabilities.
\SimName can simulate an RTL design with one top-level (at the testbench) clock and an arbitrary number of gated or divided ones.\footnote{Other work handle a single clock without any driven ones~\cite{repcut,manticore_arxiv,ash}.}

\SimName supports only a Verilog test driver, whereas Verilator allows both C++ and Verilog drivers.
We do this for pragmatic reasons: host interactions are costly on the IPU, and a C++ testbench interacts at every simulated cycle, which would be impractically slow on the IPU.\footnote{\SimName has experimental support for DPI to interface with C++.}
\SimName may fail to compile very large design whose code and state exceed the on-chip memory capacity of an IPU board ($\approx$4$\times$900~MiB).
Verilator could perhaps compile and run such massive designs, albeit, very slowly.
\SimName's philosophy is to scale out and use more IPUs for larger designs.
That said, this work explored this path only up to 4 IPUs.
Scaling simulation to 16, 64, or even 256 IPUs (225 GiB of SRAM) is left for future work.
Additionally, a single IPU tile has about 400 KiB of data memory.
So, if a design contains a single Verilog array larger than this amount, compilation fails.
Such large Verilog arrays are unlikely to appear in reasonably real silicon since large SRAMs in RTL designs (e.g., caches) are banked into much smaller arrays (e.g., 64-KiB banks).
However, these arrays might exist in non-synthesizable test benches.
Currently, users have to manually split these large arrays into smaller ones.
Finally, \SimName's compilation fails if the design has a combinational loop.

\section{Evaluation}\label{sec:evaluation}

We use the following benchmarks to evaluate \SimName:
\begin{itemize}[leftmargin=*]
    \item \bmark{mc}~\cite{DBLP:conf/fpt/TianB08} is stock option price predictor.
    \item \bmark{vta}~\cite{DBLP:journals/micro/MoreauCVRYZFJCG19} is an ML accelerator.
          We configure \bmark{vta} with \textcode{BlockIn/Out=64} (larger than the default FPGA configuration) to expose more parallelism.
    \item \smallrocket{N} is a $\text{N}\times\text{N}$ Constellation~\cite{constellation} mesh NoC consisting of $\text{N}\times\text{N} - 3$ small Rocket cores~\cite{rocket_chip} (64-bit, no FPU, no VM), generated by the Chipyard~\cite{DBLP:journals/micro/AmidBGGKLMMOPRS20} SoC generator (3 nodes connect to uncore).
          We changed \textcode{N} from 2 to 15.
    \item \largerocket{N} is similar to \smallrocket{N}, but we use large cores with an FPU and VM.
          We changed \textcode{N} from 2 to 10.
\end{itemize}

Note that \smallrocket{N} differs from \bmark{rocket} in~\secRef{bsp_t_comp} as the latter is bus-based, whereas \smallrocket{N} uses a NoC.
These benchmarks resemble contemporary chip designs, including accelerators and multicore systems.
Varying the mesh size in \smallrocket{N} and \largerocket{N} explores \SimName's performance on larger chips.
Using a generic gate library, we estimated \smallrocket{2} and \largerocket{2} to have about 200 and 320 thousand gates, respectively, while \smallrocket{15} and \largerocket{10} have about 20 million gates (excluding SRAMs).
Due to a bug in \emph{popc}, we could not evaluate BOOM~\cite{zhaosonicboom}.

We wrap all benchmarks with simple Verilog drivers, without DPI calls.\footnote{PLI calls such as \textcode{\$readmemh}, \textcode{\$display}, \textcode{\$plusargs} still exist.}
Chipyard's default simulation flow heavily uses  DPI calls to connect the simulation to services provided by a software RISC-V front-end server.
By avoiding non-RTL software, we ensure that our evaluation of both \SimName and Verilator does not include extraneous performance influences from the simulator and front-end communications.

\shortparagraph{Baseline}
Parallel Verilator is our baseline.
Other possible baselines are research artifacts that also explore parallelism.
Verilog is a complex language, so academic works, including ours, make concessions and focus on techniques rather than full language coverage and robust implementation.
These systems, unfortunately, cannot run all benchmarks (\secRef{related}), so using Verilator permits a fuller evaluation.

\shortparagraph{Evaluation Setup}
\tabRef{hw_setup} summarizes the hardware for our evaluation.
For Verilator, we use two modern data center computers:
\mAMD is the latest generation AMD server with large caches and a high core count.
The 64 cores in one socket are constructed from \emph{chiplets}~\cite{zen4} containing 8 cores.
\mIntel is a recent Intel server with no chiplets, less cache, and fewer cores.
For \SimName, we use a 4-IPU M2000 server%
\footnote{
    The M2000 is not the fastest IPU machine available.
    A newer BOW-2000 IPU clocks at 1.85~GHz (a $37\%$ increase) with the same tile count~\cite{gcore_ipu}.
}%
.

We use Verilator v5.006 (\SimName is forked from this version) with all optimizations enabled (\textcode{-O3}).
To find each design's best simulation performance on \SimName, we consider 1472, 2944, 4416, and 5888 tiles (1, 2, 3, and 4 IPUs, respectively).
On Verilator, we measured each design from 2 to 32 threads (step size of 2) because Verilator takes a long time to generate multi-threaded code for the larger designs.
\tabRef{hw_setup} reports the compilation time and memory usage.

\begin{table}
    \centering
    \resizebox{1\columnwidth}{!}{%
        \setlength{\tabcolsep}{4pt}
        \begin{tabular}{ l l l l l l l}
            \toprule
            {\bf Compiler}                         & {\bf Name/Short}         & {\bf Cores} & {\bf GHz} & {\bf MiB}   & {\bf $\times$} & {\bf Date}                                      \\
            \midrule
            \multirow{2}{*}{\small{\bf Verilator}} & {EPYC 9554 / \mAMD}      & {64}        & {3.75}    & {2/128/256} & {2}            & {Q4 2022}                                       \\
                                                   & {Xeon 6348 / \mIntel}    & {28}        & {3.5}     & {2.2/35/42} & {2}            & {Q2 2021}                                       \\
            \cmidrule(l{4pt}r{4pt}){1-7}
            {\small{\bf \SimName}}                 & {M2000 / \textcode{ipu}} & {1472}      & {1.35}    & {897}       & {4}            & {Q3 2020}                                       \\
            \midrule
            \multicolumn{7}{c}{Ubuntu 20.04 \xspace\xspace\xspace\xspace popc 3.3 (clang 16.0.0)  \xspace\xspace\xspace\xspace Verilator v5.006 \xspace\xspace\xspace\xspace g++ 10.5.0} \\
            \multicolumn{7}{c}{\SimName: tiles up to 5888\xspace\xspace\xspace\xspace\xspace\xspace Verilator: threads up to 32}                                                         \\
            \bottomrule
            \multicolumn{7}{c}{}                                                                                                                                                         \\
        \end{tabular}
    }
    \\
    \resizebox{0.75\columnwidth}{!}{%
        \begin{tabular}{l c c}
            \multicolumn{3}{c}{\textbf{Compile on Intel Xeon 6132 1.5~TiB Memory}}                                        \\
            \toprule
            {\bf min / max}    & \multicolumn{1}{c}{\bf Compile time}     & \multicolumn{1}{c}{\bf Memory usage}          \\
            \midrule
            \textbf{\SimName}  & {26s / \textcolor{ForestGreen}{\bf 40m}} & {335~MiB / 55~GiB}                            \\
            \textbf{Verilator} & {3s / \textcolor{BrickRed}{\bf 8h}}      & {223MiB / \textcolor{BrickRed}{\bf 1043~GiB}} \\
            \bottomrule
        \end{tabular}
    }
    \caption{Evaluation setup:
    {\bf Cores} is the physical core count per socket.
    \textbf{MiB} is the cache capacity (L1/L2/L3) for x64 and the on-chip memory for the IPU.
    \textbf{$\times$} is the number of sockets.
    We use \textbf{Short} names for brevity.
    We also report the min. and max. compilation time and compiler memory usage.}
    \label{tab:hw_setup}
\end{table}

\begin{figure*}
    \centering
    \includegraphics[width=\textwidth]{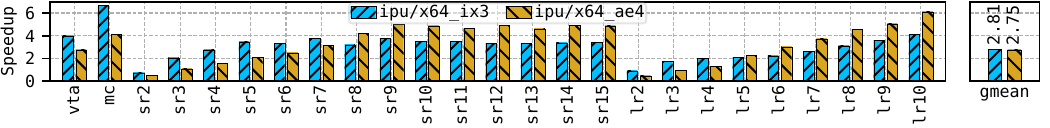}
    \caption{IPU's speedup versus multi-thread Verilator.}
    \label{fig:speedup_barplot}
\end{figure*}

\subsection{\SimName Vs. Verilator}\label{sec:superlinear}

\figRef{speedup_barplot} reports \SimName's speedup compared with Verilator.
Overall, \SimName outperforms Verilator.
The geometric mean speedups are $2.81$ and $2.75$ over \mIntel and \mAMD.
\tabRef{perf_smackdown} details the performance of each platform.
We also report size metrics for each benchmark: number of data dependence graph nodes, \fibers, x64 instructions to simulate one RTL cycle on a single thread, and Verilator's code footprint.
Furthermore, for multi-IPU points, we report the KiB size of the variables exchanged (actual exchange volume is higher due to fanout).

\subsection{Verilator's Performance}

\begin{figure}
    \subfloat[Verilator's speedup diminishes quickly for smaller designs as synchronization is costly.] {
        \label{fig:verilator_sync_cost_is_high}
        \includegraphics[width=\columnwidth]{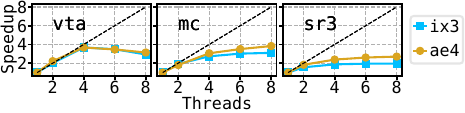}
    }\\
    \subfloat[Non-uniform communication (crossing chiplets or sockets) reduces speedups.] {
        \label{fig:verilator_comm_cost_is_high}
        \includegraphics[width=\columnwidth]{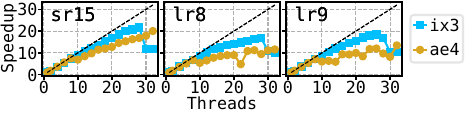}

    }\\
    \subfloat[\mAMD and \mIntel have different scaling profiles due to implementation differences.]{
        \label{fig:verilator_arch_matters}
        \includegraphics[width=\columnwidth]{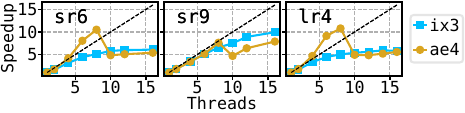}
    }
    \caption{Verilator's performance and scalability.}

\end{figure}

\tabRef{perf_smackdown} reports Verilator's best speedup relative to itself.
Verilator benefits from parallelism when a design is large (up to $22\times$ speedup).
A few points are worth considering:

\shortparagraph{Synchronization}
\figRef{verilator_sync_cost_is_high} shows that smaller designs see limited speedups.
Per~\secRef{bsp_t_sync}, we expected this behavior: synchronization cost outweighs the gains of parallelism.

\begin{figure}
    \subfloat[Simulation speed on one IPU. We start at 184 tiles ($1/8$ of an IPU) and scale to a full IPU (1472 tiles).] {
        \label{fig:ipu_is_monotonic}
        \includegraphics[width=\columnwidth]{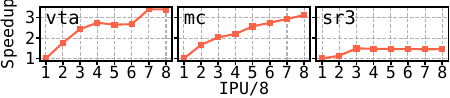}
    }\\
    \subfloat[Breakdown of simulation time.] {
        \label{fig:ipu_is_monotonic_cycles_breakdown}
        \includegraphics[width=\columnwidth]{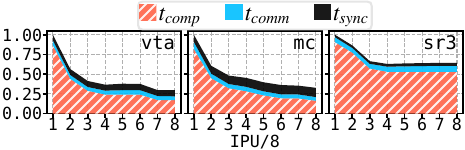}
    }
    \caption{Single-IPU speedup and simulation time breakdown.}
    \label{fig:single_ipu_scaling}

\end{figure}

\begin{figure}
    \centering
    \includegraphics[width=\columnwidth]{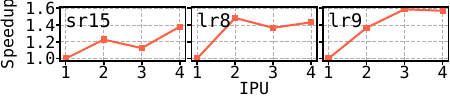}
    \caption{Performance scaling across multiple IPUs.}
    \label{fig:muli_ipu_scaling}
\end{figure}

\shortparagraph{Communication is non-uniform}
From~\secRef{bsp_t_sync} and~\figRef{rate_prngs}, we see that synchronization does not affect large designs.
\figRef{verilator_comm_cost_is_high} shows Verilator achieves significant speedups for large designs.
However, on \mAMD, speedups fade after 8 threads (chiplet boundary).
On \mIntel, we see a significant drop after 28 threads (socket boundary).
The increased communication latency across chip boundaries has a noticeable performance cost, and parallel simulators should be aware of it.

\shortparagraph{Architecture matters}
\figRef{verilator_arch_matters} shows no clear advantage between the two x64 machines.
In general, \mAMD wins for smaller designs and \mIntel for large ones.
In some cases, \mAMD shows superlinear improvement up to the chiplet boundary.
Such gains are exciting but not uncommon in RTL simulation.
Increasing cores means each core runs less code and accesses less data, reducing pressure on the local cache, which reduces cache misses and provides a performance bonus~\cite{repcut,manticore_arxiv,DBLP:conf/dac/BeamerD20,DBLP:journals/micro/Beamer20, DBLP:conf/date/LopezParadisLAWMB23}.
However, the superlinear gains disappear when the local caches cannot hold the working set or chip-cross costs diminish the value of increased parallelism.

\subsection{\SimName's Performance}

On x64, synchronization and communication were the two causes for reduced performance.
On the IPU, only off-chip communication is a bottleneck, similar to the x64's off-chiplet or -socket communication.

\shortparagraph{Single-IPU scaling}
On x64, we cannot consistently use \emph{all} cores to increase simulation speed since the synchronization or communication costs may limit parallelism gains.
On a single IPU, performance monotonically increases with additional tiles.
\figRef{ipu_is_monotonic} shows the rate for three designs as a function of the fraction of the IPU.
The IPU's limited local memory cannot fit a moderately large design on a single tile.
So, we cannot compute speedup relative to a single tile.
We use 184 tiles ($\frac{1}{8}$ of an IPU) as the baseline.
This is a fundamentally different starting point than a single thread on x64, as it already uses significant parallelism.
However, we still see improvements.\footnote{
    \figRef{ipu_is_monotonic} shows that \bmark{vta}'s performance remains flat between $\frac{4}{8}$ and $\frac{6}{8}$ of the IPU.
 Such \emph{staircase} behavior is a characteristic of highly regular fine-grain parallelism where there are many equal-size straggler \threads.
 Let us explain using a hypothetical but similar example with 12 equal-size \fibers $f_1$ until $f_{12}$.
 Suppose at first, we run these \fibers on 4 balanced \threads:
    $p_1 = \lbrace f_1, f_2, f_3 \rbrace$, ..., $p_4 = \lbrace f_{10}, f_{11}, f_{12} \rbrace$.
 Using 5 balanced {\threads}, we will have:
    $p_1^{\prime} = \lbrace f_1, f_2, f_3 \rbrace$, ..., $p_3^\prime = \lbrace f_7, f_8 \rbrace$, $p_4^\prime = \lbrace f_{9}, f_{10} \rbrace$, $p_5^\prime = \lbrace f_{11}, f_{12} \rbrace$.
 In other words, increasing the number of \threads did not reduce the worst-case execution time.
 However, with 6 \threads, we will have:
    $p_1^{\prime\prime} = \lbrace f_1, f_2 \rbrace$, ..., $p_6{\prime\prime} = \lbrace f_{11}, f_{12} \rbrace$, i.e., a 33\% improvement in execution time.
}

\figRef{ipu_is_monotonic_cycles_breakdown} shows the breakdown of simulation time for each design.
The vertical axis is normalized to $\frac{1}{8}$ of the IPU.
Communication ($t_{comm}$) and synchronization ($t_{sync}$) remain roughly constant while computation time ($t_{comp}$) decreases with additional tiles.
However, in~\smallrocket{3}, the improvement in $t_{comp}$ ends due to \fiber imbalance (see~\secRef{bsp_t_comp}).
The IPU performance is non-decreasing because of its low-cost communication and synchronization, but, to achieve the best performance on x64, a hardware developer must select the parallelism for each design \emph{and} machine (\mIntel or \mAMD).

\shortparagraph{Multi-IPU scaling}
Within one IPU, communication is relatively cheap, which facilitates scaling.
However, communication and synchronization across IPU boundaries is expensive (\secRef{bsp_t_comm}).
Therefore, preserving performance monotonicity across IPUs is challenging: crossing IPUs is similar to crossing chiplets or sockets since off-chip communication latency increases abruptly (see~\figRef{ipu_bw}).
However, non-uniform communication emerges much later on the IPU (after 1472 tiles rather than 8 or 28 threads).
\figRef{muli_ipu_scaling} shows the simulation speed across multiple IPUs.
Even for very large designs, running at maximum parallelism may yield a poorer result so that fewer IPUs can produce marginal gains in some cases.

Improvements are also much smaller off-chip.
Going from 1472 tiles to 5888 tiles ($4\times$) in \largerocket{9} improves performance by $60\%$.
However, a $60\%$ gain is still attractive: on x64, it is difficult to scale beyond 28 threads (\mIntel), but on the IPU, we increase performance to 5888 tiles ($210\times$ more parallel).

\begin{figure}
    \centering
    \includegraphics[width=\columnwidth]{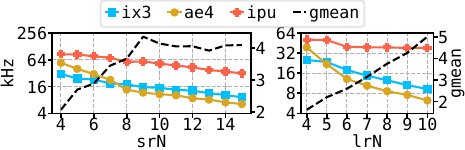}
    \caption{Coping with increasing design size.
        The left axis shows the best simulation rate.
        The right axis shows the geomean speedup of \SimName against Verilator (dashed lines).}
    \label{fig:ipu_vs_verilator}
\end{figure}

\shortparagraph{Performance resilience}
\SimName can strongly  scale the simulation rate within and across IPUs.
But can we can maintain a constant simulation rate as we scale the design size (weak scaling)?
\figRef{ipu_vs_verilator} shows the maximum simulation rate of \SimName and Verilator as a function of mesh size in \smallrocket{N} and \largerocket{N}.
Neither \SimName nor Verilator can keep the simulation rate perfectly constant, but \SimName is better.
For instance, at the right of~\figRef{ipu_vs_verilator}, there is a long period in which \SimName simulates larger designs at the same rate.
Verilator's rate slowly drops in this region, and the speedup (\SimName vs. Verilator) increases.

\begin{figure}
    \centering
    \includegraphics[width=\columnwidth]{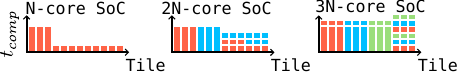}
    \caption{\Fiber imbalance allows us to keep the simulation rate constant despite increasing design size.}
    \label{fig:weak_scaling}
\end{figure}

While \fiber imbalance severely limits the performance of a small or medium RTL design, limiting strong scaling, it actually \emph{enables} better weak scaling.
\figRef{weak_scaling} shows how it happens.
Consider an SoC with $N$ cores and a sizable imbalance among its \fibers, as shown on the left.
If we double the size of the SoC, we double the number of \fibers.
Because only a small portion of \fibers have significantly longer execution time, we can tolerate increasingly larger designs and keep the simulation rate constant using unused parallel resources (\figRef{weak_scaling}).
However, the utilization of tiles starts to balance at some point, after which increasing the design size decreases the simulation rate.

\begin{table*}
    \centering
    \resizebox{1\textwidth}{!}{%
        \setlength{\tabcolsep}{3pt}
        \begin{tabular}{ l lllllllllllllllllll}
\toprule
\multirow{2}{*}{\rotatebox{45}{\textbf{Bench}}}             & \multicolumn{4}{c}{\textbf{\mIntel}} & \multicolumn{4}{c}{\textbf{\mAMD}} & \multicolumn{2}{c}{\textbf{\SimName}} & \multicolumn{3}{c}{\textbf{Speedup}} & \multicolumn{4}{c}{\textbf{}} \\
\cmidrule(l{4pt}r{4pt}){2-5}
\cmidrule(l{4pt}r{4pt}){6-9}
\cmidrule(l{4pt}r{4pt}){10-11}
\cmidrule(l{4pt}r{4pt}){12-14}
{}                   & \textbf{st-kHz}      & \textbf{mt-kHz}      & \textbf{\#T}         & \textbf{gain}        & \textbf{st-kHz}      & \textbf{mt-kHz}      & \textbf{\#T}         & \textbf{gain}        & \textbf{kHz}         & \textbf{\#T}         & \textbf{\mIntel}     & \textbf{\mAMD}       & \textbf{gmean}       & \textbf{MiB}         & \textbf{\#I (M)}     & \textbf{\#N (K)}     & \textbf{\#F (K)}     & \textbf{Int.}        & \textbf{Ext.}        \\
\midrule
\textbf{vta}         & {30.91}              & {113.75}             & {4}                  & {3.7}                & \textcolor{RoyalBlue}{44.79} & \textcolor{RoyalBlue}{164.73} & {4}                  & {3.7}                & {454.10}             & {1472}               & \textcolor{ForestGreen}{3.99} & \textcolor{ForestGreen}{2.76} & \textcolor{ForestGreen}{3.32} & {1.5}                & {  0.17}             & { 23.5}              & { 6.0}               & { 28.7}              & {---}                \\
\textbf{mc}          & {28.68}              & {88.96}              & {8}                  & {3.1}                & \textcolor{RoyalBlue}{37.55} & \textcolor{RoyalBlue}{143.88} & {8}                  & {3.8}                & {592.83}             & {1472}               & \textcolor{ForestGreen}{6.66} & \textcolor{ForestGreen}{4.12} & \textcolor{ForestGreen}{5.24} & {1.0}                & {  0.15}             & { 26.9}              & { 7.5}               & { 24.2}              & {---}                \\
\textbf{sr2}         & {123.40}             & {76.22}              & {2}                  & {0.6}                & \textcolor{RoyalBlue}{176.49} & \textcolor{RoyalBlue}{145.75} & {4}                  & {0.8}                & {91.20}              & {1472}               & \textcolor{BrickRed}{0.74} & \textcolor{BrickRed}{0.52} & \textcolor{BrickRed}{0.62} & {1.2}                & {  0.06}             & { 12.7}              & { 2.8}               & { 12.8}              & {---}                \\
\textbf{sr3}         & {20.95}              & {40.95}              & {8}                  & {2.0}                & \textcolor{RoyalBlue}{28.71} & \textcolor{RoyalBlue}{77.66} & {8}                  & {2.7}                & {83.95}              & {1472}               & \textcolor{ForestGreen}{2.05} & {1.08}               & {1.49}               & {3.1}                & {  0.17}             & { 36.3}              & { 8.1}               & { 33.9}              & {---}                \\
\textbf{sr4}         & \textcolor{RoyalBlue}{8.79} & {30.93}              & {22}                 & {3.5}                & {7.23}               & \textcolor{RoyalBlue}{54.79} & {8}                  & {7.6}                & {85.09}              & {1472}               & \textcolor{ForestGreen}{2.75} & {1.55}               & \textcolor{ForestGreen}{2.07} & {5.5}                & {  0.32}             & { 68.2}              & { 15.3}              & { 63.5}              & {---}                \\
\textbf{sr5}         & \textcolor{RoyalBlue}{5.23} & {24.34}              & {26}                 & {4.6}                & {4.26}               & \textcolor{RoyalBlue}{40.09} & {8}                  & \underline{9.4}      & {84.28}              & {1472}               & \textcolor{ForestGreen}{3.46} & \textcolor{ForestGreen}{2.10} & \textcolor{ForestGreen}{2.70} & {8.6}                & {  0.50}             & { 107.9}             & { 24.2}              & { 101.5}             & {---}                \\
\textbf{sr6}         & \textcolor{RoyalBlue}{3.53} & {23.11}              & {20}                 & {6.5}                & {2.90}               & \textcolor{RoyalBlue}{30.51} & {8}                  & \underline{10.5}     & {76.63}              & {1472}               & \textcolor{ForestGreen}{3.32} & \textcolor{ForestGreen}{2.51} & \textcolor{ForestGreen}{2.89} & {12.2}               & {  0.72}             & { 156.0}             & { 35.0}              & { 145.4}             & {---}                \\
\textbf{sr7}         & \textcolor{RoyalBlue}{2.47} & {18.83}              & {28}                 & {7.6}                & {2.10}               & \textcolor{RoyalBlue}{22.72} & {8}                  & \underline{10.8}     & {71.33}              & {2944}               & \textcolor{ForestGreen}{3.79} & \textcolor{ForestGreen}{3.14} & \textcolor{ForestGreen}{3.45} & {16.6}               & {  0.99}             & { 212.6}             & { 47.7}              & { 199.2}             & { 0.9}               \\
\textbf{sr8}         & \textcolor{RoyalBlue}{1.82} & \textcolor{RoyalBlue}{17.94} & {26}                 & {9.9}                & {1.58}               & {13.66}              & {8}                  & \underline{8.7}      & {57.39}              & {2944}               & \textcolor{ForestGreen}{3.20} & \textcolor{ForestGreen}{4.20} & \textcolor{ForestGreen}{3.66} & {21.6}               & {  1.29}             & { 277.3}             & { 62.3}              & { 259.0}             & { 1.1}               \\
\textbf{sr9}         & \textcolor{RoyalBlue}{1.37} & \textcolor{RoyalBlue}{15.56} & {28}                 & {11.4}               & {1.22}               & {11.72}              & {32}                 & {9.6}                & {58.79}              & {4416}               & \textcolor{ForestGreen}{3.78} & \textcolor{ForestGreen}{5.02} & \textcolor{ForestGreen}{4.35} & {27.4}               & {  1.65}             & { 351.4}             & { 78.8}              & { 328.8}             & { 1.7}               \\
\textbf{sr10}        & \textcolor{RoyalBlue}{1.06} & \textcolor{RoyalBlue}{15.03} & {24}                 & {14.1}               & {0.97}               & {10.83}              & {32}                 & {11.1}               & {52.77}              & {2944}               & \textcolor{ForestGreen}{3.51} & \textcolor{ForestGreen}{4.87} & \textcolor{ForestGreen}{4.14} & {33.8}               & {  2.03}             & { 433.5}             & { 97.2}              & { 396.3}             & { 1.3}               \\
\textbf{sr11}        & \textcolor{RoyalBlue}{0.85} & \textcolor{RoyalBlue}{13.59} & {26}                 & {16.0}               & {0.79}               & {10.21}              & {32}                 & {12.9}               & {47.71}              & {5888}               & \textcolor{ForestGreen}{3.51} & \textcolor{ForestGreen}{4.67} & \textcolor{ForestGreen}{4.05} & {40.9}               & {  2.47}             & { 524.6}             & { 117.5}             & { 488.0}             & { 3.3}               \\
\textbf{sr12}        & \textcolor{RoyalBlue}{0.70} & \textcolor{RoyalBlue}{12.98} & {28}                 & {18.5}               & {0.65}               & {8.79}               & {32}                 & {13.5}               & {43.30}              & {5888}               & \textcolor{ForestGreen}{3.34} & \textcolor{ForestGreen}{4.93} & \textcolor{ForestGreen}{4.05} & {48.6}               & {  2.93}             & { 623.7}             & { 139.7}             & { 579.5}             & { 3.1}               \\
\textbf{sr13}        & \textcolor{RoyalBlue}{0.58} & \textcolor{RoyalBlue}{11.40} & {28}                 & {19.5}               & {0.54}               & {8.18}               & {32}                 & {15.2}               & {37.83}              & {4416}               & \textcolor{ForestGreen}{3.32} & \textcolor{ForestGreen}{4.62} & \textcolor{ForestGreen}{3.92} & {56.9}               & {  3.44}             & { 731.1}             & { 163.9}             & { 665.7}             & { 2.3}               \\
\textbf{sr14}        & \textcolor{RoyalBlue}{0.50} & \textcolor{RoyalBlue}{10.37} & {28}                 & {20.8}               & {0.44}               & {7.09}               & {32}                 & {16.1}               & {34.98}              & {5888}               & \textcolor{ForestGreen}{3.37} & \textcolor{ForestGreen}{4.93} & \textcolor{ForestGreen}{4.08} & {65.9}               & {  3.99}             & { 847.0}             & { 189.9}             & { 775.2}             & { 3.4}               \\
\textbf{sr15}        & \textcolor{RoyalBlue}{0.43} & \textcolor{RoyalBlue}{9.22} & {28}                 & {21.6}               & {0.33}               & {6.51}               & {32}                 & {20.0}               & {31.69}              & {5888}               & \textcolor{ForestGreen}{3.44} & \textcolor{ForestGreen}{4.86} & \textcolor{ForestGreen}{4.09} & {75.6}               & {  4.58}             & { 972.2}             & { 217.9}             & { 886.9}             & { 3.7}               \\
\textbf{lr2}         & {69.07}              & {70.69}              & {2}                  & {1.0}                & \textcolor{RoyalBlue}{123.55} & \textcolor{RoyalBlue}{132.09} & {8}                  & {1.1}                & {64.58}              & {1472}               & \textcolor{BrickRed}{0.91} & \textcolor{BrickRed}{0.49} & \textcolor{BrickRed}{0.67} & {1.6}                & {  0.09}             & { 16.5}              & { 3.7}               & { 16.2}              & {---}                \\
\textbf{lr3}         & \textcolor{RoyalBlue}{8.74} & {33.89}              & {12}                 & {3.9}                & {7.79}               & \textcolor{RoyalBlue}{60.93} & {8}                  & {7.8}                & {58.73}              & {1472}               & {1.73}               & \textcolor{BrickRed}{0.96} & {1.29}               & {5.7}                & {  0.36}             & { 59.4}              & { 13.3}              & { 55.5}              & {---}                \\
\textbf{lr4}         & \textcolor{RoyalBlue}{4.13} & {25.27}              & {22}                 & {6.1}                & {3.61}               & \textcolor{RoyalBlue}{38.97} & {8}                  & \underline{10.8}     & {50.93}              & {5888}               & \textcolor{ForestGreen}{2.02} & {1.31}               & {1.62}               & {11.1}               & {  0.73}             & { 118.2}             & { 26.7}              & { 109.9}             & { 1.8}               \\
\textbf{lr5}         & \textcolor{RoyalBlue}{2.36} & \textcolor{RoyalBlue}{23.56} & {26}                 & {10.0}               & {2.15}               & {21.87}              & {8}                  & \underline{10.2}     & {50.09}              & {5888}               & \textcolor{ForestGreen}{2.13} & \textcolor{ForestGreen}{2.29} & \textcolor{ForestGreen}{2.21} & {17.8}               & {  1.20}             & { 192.4}             & { 43.4}              & { 178.4}             & { 2.0}               \\
\textbf{lr6}         & \textcolor{RoyalBlue}{1.50} & \textcolor{RoyalBlue}{17.86} & {28}                 & {11.9}               & {1.43}               & {13.15}              & {30}                 & {9.2}                & {39.84}              & {1472}               & \textcolor{ForestGreen}{2.23} & \textcolor{ForestGreen}{3.03} & \textcolor{ForestGreen}{2.60} & {26.0}               & {  1.77}             & { 282.8}             & { 63.7}              & { 256.2}             & {---}                \\
\textbf{lr7}         & \textcolor{RoyalBlue}{1.03} & \textcolor{RoyalBlue}{14.73} & {28}                 & {14.3}               & {1.01}               & {10.41}              & {30}                 & {10.3}               & {39.00}              & {2944}               & \textcolor{ForestGreen}{2.65} & \textcolor{ForestGreen}{3.74} & \textcolor{ForestGreen}{3.15} & {35.8}               & {  2.45}             & { 389.4}             & { 87.7}              & { 354.8}             & { 1.3}               \\
\textbf{lr8}         & \textcolor{RoyalBlue}{0.74} & \textcolor{RoyalBlue}{12.52} & {28}                 & {16.9}               & {0.74}               & {8.60}               & {32}                 & {11.6}               & {39.02}              & {2944}               & \textcolor{ForestGreen}{3.12} & \textcolor{ForestGreen}{4.54} & \textcolor{ForestGreen}{3.76} & {47.0}               & {  3.24}             & { 511.8}             & { 115.4}             & { 463.9}             & { 1.0}               \\
\textbf{lr9}         & \textcolor{RoyalBlue}{0.58} & \textcolor{RoyalBlue}{10.63} & {26}                 & {18.5}               & {0.56}               & {7.57}               & {32}                 & {13.4}               & {38.22}              & {4416}               & \textcolor{ForestGreen}{3.60} & \textcolor{ForestGreen}{5.05} & \textcolor{ForestGreen}{4.26} & {59.8}               & {  4.14}             & { 651.3}             & { 146.7}             & { 595.8}             & { 1.6}               \\
\textbf{lr10}        & \textcolor{RoyalBlue}{0.45} & \textcolor{RoyalBlue}{9.27} & {28}                 & {20.6}               & {0.37}               & {6.27}               & {32}                 & {17.0}               & {38.24}              & {5888}               & \textcolor{ForestGreen}{4.13} & \textcolor{ForestGreen}{6.10} & \textcolor{ForestGreen}{5.02} & {74.0}               & {  5.12}             & { 806.4}             & { 181.7}             & { 734.9}             & { 3.2}               \\
\midrule
            &  &  &  &  &  &  &  &  &   \multicolumn{2}{c}{\textbf{gmean}}       & {2.81}               & {2.75}               & {2.78}               &  &  &  &  &  \\
\bottomrule
\end{tabular}

    }
    \caption{%
        \textbf{st-kHz}, \textbf{mt-kHz} are single- and multi-thread Verilator performance (\textcolor{RoyalBlue}{blue} is best of \mIntel--\mAMD).
        \textbf{kHz} is best \SimName rate.
        \textbf{gain} is Verilator's self-relative speedup (underscored \underline{superlinear}).
        \textbf{\#T} is threads or tile count.
        \textbf{Speedup} is \SimName vs. Verilator (\textcolor{ForestGreen}{green} $\geq 2$ and \textcolor{BrickRed}{red} $<1$).
        \textbf{gmean} is reported across machines and benchmarks.
        \textbf{MiB} is Verilator's binary size.
        \textbf{\#I} is the millions of x64 instructions per RTL cycle (Verilator).
        \textbf{\#N} is thousands of data dependence graph nodes.
        \textbf{\#F} is thousands of \fibers.
        \textbf{Int.} and \textbf{Ext.} are KiBs on- and off-chip cut size (lower than actual communication volume due to fanout).
    }
    \label{tab:perf_smackdown}
\end{table*}

\subsection{Cost Comparison}

The IPU's performance advantage for large designs makes it more cost-effective than other systems.
The cloud hosting service GCore offered \textcode{IPU-POD4} classic instances (an M2000, see~\tabRef{hw_setup}) for \$2.13 per hour~\cite{gcore_ipu}.
A \textcode{Dv4} Microsoft Azure instance (Xeon 8272CL) costs \$0.048 per hour per core~\cite{azure_instance_prices} or \$0.77 per hour for 16 cores.
We use the \smallrocket{15} design to briefly compare the cost of running long and short simulations on \SimName and Verilator.
We exclude compilation time and cost from our analysis.

\shortparagraph{Single Long Test}
Consider simulating \smallrocket{15} for 1 billion cycles.
On \textcode{Dv4}, the simulation scales from 222 Hz (1 thread) to 4.88 kHz (16 threads, a superlinear 22$\times$ speedup, but slows down beyond 16 threads).
The rate on \textcode{IPU-POD4} scales from 22.94 kHz (1 IPU) to 31.69 kHz (4 IPUs).
So, \textcode{IPU-POD4} finishes the simulation in 9 hours, costing \$19.20.
But, \textcode{Dv4} takes 57 hours and costs \$43.78 (16 threads).

A back-of-the-envelope calculation shows that \textcode{IPU-POD4} is always more cost-effective than \textcode{Dv4} irrespective of the number of rented cores.
Let $t$ be the number of Verilator threads and $s$ be its speedup (vs. single-thread performance).
Four IPUs run 142.74$\times$ faster than single-thread \textcode{Dv4}.
So long as $\frac{s}{t} < 142.74 \times \frac{0.048}{2.13} = 3.2$, \textcode{Dv4} with $t$ threads will cost more than \textcode{IPU-POD4}.
Since linear scaling is $\frac{s}{t} = 1$, Verilator would become cost-effective only at a $3.2\times$ superlinear scaling, which is very far from what we observe.

\begin{figure}
    \centering
    \includegraphics[width=\columnwidth]{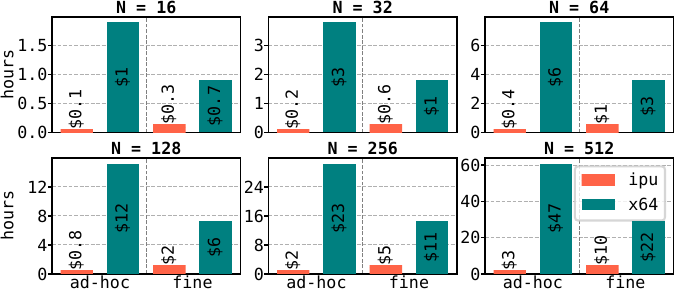}
    \caption{Nightly test simulation time (in hours, \smallrocket{15}) on a 16-core \textcode{Dv4} instance (x64) and the \textcode{IPU-POD4} classic (\textcode{ipu}).
    Numbers on bars are total cost and \textcode{N} is number of run tests.}
    \label{fig:price_compare}

\end{figure}

\shortparagraph{Several Short Tests}
We now run many short---1 million cycles---``nightly regression'' tests using two strategies.
First, the most straightforward strategy:
On \textcode{Dv4}, we assign a core per test, running 16 tests in parallel.
Since it is impossible to assign one IPU tile to each test (there is not enough memory), we conservatively assign one IPU to each test and run four tests in parallel at a time on \textcode{IPU-POD4}.
We call this ad-hoc parallelism.
Second, we run each test with an optimal number of threads or tiles, which is 16 on~\textcode{Dv4} and 5888 on \textcode{IPU-POD4}.
We call this fine-grained parallelism, as for this benchmark, we end up exploiting parallelism within each test but running the tests one after the other.

Admittedly, these scheduling strategies do not explore the performance-cost space systematically; they illustrate the two obvious alternatives.
\figRef{price_compare} shows the time to finish \textcode{N} tests.
The numbers on each bar show the total cost.
On the IPU, ad-hoc parallelism is more cost-effective because fine-grained parallelism scales sublinearly.
On the x64, we see the opposite trend since ad-hoc parallelism scales almost linearly, while fine-grained parallelism scales superlinearly (22$\times$ on 16 threads).
The fine-grained parallelism finishes faster.
Regardless, \SimName is cheaper for both approaches since its performance edge exceeds the cost difference.

\shortparagraph{Power and Energy Estimates}
We measured that the 4 IPUs consume 185W.
Due to security limitations, we could not measure power draw on our x64 baselines.
The \mAMD and \mIntel TDP's are 320W and 235W, respectively.
We estimate the power draw to be 80W and 118W as we use a quarter of the cores on \mAMD and half on \mIntel.
\smallrocket{15} on the IPU is 4.86$\times$ faster than \mAMD and 3.44$\times$ faster than \mIntel, so the IPU's energy draw is about 2$\times$ lower than x64.

\tb{I think we should not put the price once rounded per hours. Some cloud providers bill per minutes and even per seconds. }

\subsection{Comparison with Other Systems}
\shortparagraph{RepCut}
RepCut~\cite{repcut} is a BSP RTL simulator (full-cycle) for \textcode{firrtl}~\cite{DBLP:conf/iccad/IzraelevitzKLLW17} on x64.
RepCut demonstrates superlinear speedups within and across sockets (we saw a similar effect with chiplets).
It improved simulation up to 32 cores (48-core machine) but showed no gains beyond that.

We were unable to compare directly for frustrating practical reasons.
Chipyard's CIRCT backend has replaced the first \textcode{firrtl} Scala compiler that RepCut is based on.
The new version cannot ingest the \smallrocket{N} and \largerocket{N} designs produced by Chipyard---there is no reliable way to convert Verilog back to \textcode{firrtl}.
However, we were able to simulate an older Rocket SoC (based on git hash \href{https://github.com/chipsalliance/rocket-chip/}{\small{4276f17f9}}) with RepCut and compare it against \SimName.
We ported our lightweight Chipyard test driver, free of DPI calls, to the Rocket SoC and removed all internal print statements.
We found that the stock test driver for Rocket SoC (and Chipyard) severely limits performance.
RepCut reports a 1-core Rocket SoC simulates at $\approx$10~kHz on Verilator and $\approx$50~kHz on RepCut (single-thread).
We reproduced these results.
However, with our streamlined test driver, the benchmarks ran at 276~kHz on Verilator and 75~kHz on RepCut (on \mAMD), showing that the original Rocket test driver is a debatable baseline.

\begin{figure}
    \includegraphics[width=\columnwidth]{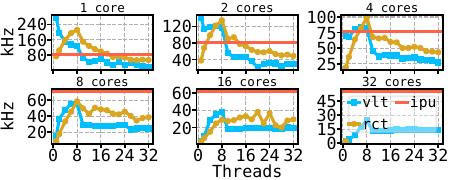}
    \caption{
        Performance of RepCut (\textcode{rct})~\cite{repcut}, Verilator (\textcode{vlt}), and \textcode{ipu}.
        Figure legend is on the bottom right.
    }
    \label{fig:vlt_rct_ipu}
\end{figure}

\figRef{vlt_rct_ipu} compares Verilator, RepCut, and \SimName for various SoC sizes.
We ran Verilator and RepCut simulations on \mAMD up to 32 threads.
\SimName ran on a single IPU.
Verilator is fastest for smaller designs, RepCut gains a small advantage for medium SoCs, and \SimName performs best for the largest.
Code generated by RepCut for the 32-core SoC crashes clang.

\begin{figure}
    \includegraphics[width=\columnwidth]{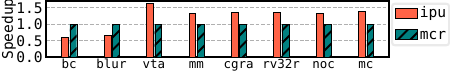}
    \caption{
        Comparison of \textcode{ipu} and Manticore (\textcode{mcr}).
    }
    \label{fig:ipu_mcr}

\end{figure}

\shortparagraph{Manticore}
Manticore~\cite{manticore_arxiv} is a 225-core, statically scheduled, deeply pipelined architecture designed for BSP RTL simulation and prototyped on an FPGA.
Manticore's compiler frontend does not support Verilog's packed arrays used abundantly in \largerocket{N} and \smallrocket{N}.
Moreover, since FPGAs have limited memory, large designs do not fit on Manticore.
\figRef{ipu_mcr} compares \SimName (1472 tiles) to Manticore (225 cores) using the raw numbers reported in their work~\cite{manticore_arxiv} (see~\cite{manticore_arxiv} for the description of the designs).
Manticore's huge register file lets it achieve a higher single-core rate than the IPU.
So, the small \bmark{bc} design runs faster on it, but the larger \bmark{vta} and \bmark{mc} designs benefit from \SimName's greater parallelism.


\subsection{Partitioning Strategies}
So far, we have used the partitioning strategy outlined in~\secRef{partitioning}.
This section considers alternative strategies for partitioning \fibers within and across IPUs.

\begin{figure}
    \centering
    \includegraphics[width=\columnwidth]{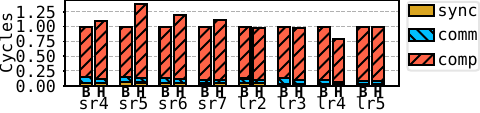}
    \caption{Comparison of \SimName's bottom-up SLB algorithm (\textbf{B}) against RepCut's hypergraph approach (\textbf{H})~\cite{repcut}.
 The vertical axis shows normalized IPU machine cycles per RTL cycle (lower is better).}
    \label{fig:merge_compare_breakdown}
\end{figure}

\shortparagraph{Single-IPU Partitioning}
RepCut~\cite{repcut} formulates SLB as a hypergraph partitioning problem where hypergraph nodes \fibers and hyperedges represent duplicated clusters across \fibers.
We implemented this strategy as an alternative.
\figRef{merge_compare_breakdown} compares the default bottom-up (\textbf{B},~\secRef{partitioning}) strategy against hypergraph partitioning (\textbf{H}) on a single IPU (1472-way partitioning).
Neither strategy is uniformly better.
Bottom-up performs best with \smallrocket{N}, whereas hypergraph is sometimes better with \largerocket{N}.

\begin{figure}
    \centering
    \includegraphics[width=\columnwidth]{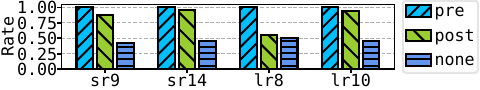}
    \caption{Normalized simulation rate for 4-IPU partitioning strategies.
 Partitioning \fibers \textcode{pre} merge performs better than partitioning \threads \textcode{post} merge.
 Ignoring the muli-IPU configuration (\textcode{none}) yields vastly inferior results.
        \me{TODO: Add another plot that show the cut size in each.}
 }
    \label{fig:device_partitioning_compare}

\end{figure}

\shortparagraph{Multi-IPU Partitioning}
\figRef{device_partitioning_compare} compares three strategies for multi-device partitioning of 4 IPUs:
(pre) partition \fibers across IPUs before merging them into \threads (default \SimName strategy), (post) partition \threads across IPU, i.e., after merging \fibers into \threads, and (none) does not partition, i.e., multi-IPU oblivious.

Not partitioning \fibers or \threads across IPUs yields inferior performance.
Partitioning \fibers works better than partitioning \threads.
The former approach offers more degrees of freedom for partitioning since the earlier \thread merge may suboptimally absorb some \emph{good cuts} and land in a region of the design space that is only locally optimal.

\subsection{Discussion}

Fast simulation requires a simulator that exploits the fine-grain parallelism of RTL and effectively utilizes the features of the underlying hardware platform.
Verilator fails to scale because its frequent fine-grain synchronization overuses the x64's costly synchronization and communication.
\SimName scales better, albeit from lower single-core performance, because an IPU efficiently supports BSP synchronization and low-latency communication.
However, the IPU's high off-chip latency demands effective RTL partitioning to minimize cross-chip traffic.
When we started this project, we expected no speed gains from multiple IPUs and only planned to use additional IPUs when we ran out of memory.
We were surprised to find speedups to even 5888 cores and believe compiler improvements would increase performance further.

VLSI design practices explain why speedups are possible on thousands of cores.
Optimizing circuit performance for synthesis, placement, and routing requires a \emph{floorplan} that is aware of physical constraints such as pin placement.
A good design facilitates floorplanning and optimizes off-chip communication in its simulation---there is a natural minimal cut.
The critical path length in a VLSI circuit limits the clock rate, just as the straggler \fiber and its cone of logic limit the simulation rate.
A fast circuit design minimizes the critical path length, indirectly minimizing the critical cone of logic (area) and producing many fibers.
In general, faster circuits should simulate faster and utilize parallelism better.

The lessons from \SimName may help apply BSP for RTL simulation on other parallel architectures.
Low-latency memory (SRAM) capacity is the main enabler of high-speed medium-to-large RTL simulations (and a bottleneck on x86).
Other architectures such as Groq~\cite{DBLP:conf/isca/AbtsRSWBHBTKKHL20, DBLP:conf/isca/AbtsKLKBBPADHTB22} or Cerebras~\cite{DBLP:journals/micro/Lauterbach21} offer considerable low-latency memory and might be good platforms for RTL simulation.
By contrast, the NVIDIA H100 GPU has only $\approx$50~MiBs of shared on-chip memory.
As a result, GPUs are unlikely to perform well using the BSP execution model.
Besides SRAM capacity, low-cost synchronization is essential for BSP RTL simulation on an accelerator.
The IPU (and perhaps a Groq-like architecture) has predictably low synchronization costs, as computation and communication are almost entirely statically scheduled.
Tentative experiments found (not presented in the paper) that full-device synchronization on GPUs does not share the same predictably low latencies.

\section{Related Work}\label{sec:related}

\SimName is the first RTL simulator for a few thousand cores.
Prior work used tens of CPU and GPU cores or a few hundred specialized cores.
In addition, prior work uses software or FPGA emulation to simulate thousand-core SoCs.

\subsection{Tens of Cores}
Verilator and RepCut~\cite{repcut} are parallel, full-cycle RTL simulators that target commodity processors with few tens of cores.
Both are limited by the x64's expensive synchronization and communication, as shown earlier.
Other research on parallel, event-driven techniques focuses on finer granularity rather than full-cycle simulation.
They employ smart concurrency techniques on x64 to avoid computation/communication~\cite{ahmad_numa,DBLP:conf/date/KimCY11,DBLP:conf/date/YangHKKCPK14,DBLP:conf/vlsid/LiGL04}.
SAGA~\cite{DBLP:conf/dac/VincoCBF12} achieves a $16\times$ parallel speedup on GPUs by statically scheduling SystemC~\cite{systemc}.
GCS~\cite{DBLP:conf/dac/ChatterjeeDB09, DBLP:conf/date/ChatterjeeDB09,DBLP:journals/todaes/ChatterjeeDB11} employs \emph{levelization}~\cite{DBLP:conf/dac/WangHPZ87,DBLP:conf/dac/WangM90} and acclerates gate-level simulation on GPUs.

\subsection{Hundreds of Cores}
100+ core general-purpose machines are uncommon, except in GPUs and accelerators.
Since RTL is irregular, SIMD execution on a GPU typically yields low thread utilization.
Qian et al. ~\cite{DBLP:conf/iccad/QianD11} describe a GPU-accelerated, event-driven simulator with a single thread per GPU core (i.e., $\frac{1}{32}$ Warp utilization).
RTLFlow~\cite{rtlflow} fully utilizes a GPU by independently simulating a single design driven by multiple test vectors.
RTLFlow performs comparably to Verilator with 1K test vectors but runs $40\times$ faster with 64K tests.
RTLFlow is limited by available GPU memory, as described in~\cite{ash}.
Nexus~\cite{nexus_emulation:22} is an FPGA-based parallel RTL simulator with a systolic array of 240 8-bit processors.
Like Manticore, Nexus suffers from limited SRAM resources on FPGAs.
ASH~\cite{ash} extends the Swarm architecture~\cite{DBLP:conf/micro/JeffreySYES15} with prioritized dataflow to accelerate RTL simulation.
It comprises 256 simple x64 cores with dedicated task queues to support efficient event-driven simulation.
ASH demonstrated a $32\times$ speedup over Verilator running on a simulator (it has not been prototyped).

\subsection{Thousands of Cores}
Most previous work on thousand-way parallel simulation used C++ processor (architectural) models, not RTL.
An RTL simulator parallelizes a model (the code), whereas an architectural simulator \emph{implements} parallel models.
Hence, the simulator developer must parallelize the code rather than our compiler.
Moreover, many architectural simulators compromise modeling accuracy to enable efficient parallel execution~\cite{DBLP:conf/isca/SanchezK13, DBLP:conf/hpca/MillerKKGBCEA10, DBLP:journals/sigarch/ChenAD09,DBLP:conf/date/ZurstrassenCJYXL23}, which is incompatible with RTL's rigid semantics.
Some architectural simulators use multiple machines for large-scale simulation~\cite{DBLP:conf/ispass/FuW14, DBLP:conf/hpca/MillerKKGBCEA10, DBLP:conf/date/LopezParadisLAWMB23, Herbst2024SwitchboardAO}.
Like \SimName, the motivation for distributed simulation is to use multiple machines' computing and memory resources.

Metro-MPI~\cite{DBLP:conf/date/LopezParadisLAWMB23} is a framework for \emph{manually} connecting independent simulations of hardware components that interact over a clearly defined interface, such as a NoC.
It does not compile an RTL design into an executable that can simulate a design.
Instead, it utilizes coarse-grain parallelism and concurrently simulates one or a few components on each core.
Theoretically, Metro-MPI could exploit fine-grain parallelism by dedicating several cores to a simulation and running Verilator in parallel.
However, realistically, it requires implementation and fine-tuning on a per-project basis.
By contrast, \SimName automatically (without developer assistance) extracts fine-grain parallelism across an entire RTL design, groups it into appropriate-sized computations, and maps them to available computing resources.
That said, Metro-MPI manages to evaluate a design that is $\approx$20$\times$ larger than the largest design we evaluated in this work (i.e., \smallrocket{15} and \largerocket{10}).
We estimate this difference: Metro-MPI evaluates a 1000-core chip containing 10 billion transistors, or perhaps 1 billion generic gates, including SRAM.
As mentioned in~\secRef{evaluation}, \largerocket{10} has $\approx$20 million gates, excluding SRAM.
Including SRAM in our estimation, \largerocket{10} probably contains 50 million gates, 20$\times$ smaller than Metro-MPI's workload.

Emulating large systems on FPGAs is an alternative to simulation, but capacity and compile time are significant challenges.
FireSim~\cite{DBLP:journals/micro/KarandikarMKBAL19} and DIABLO~\cite{DBLP:conf/asplos/TanQCAP15} simulate warehouse-scale computers, but individual FPGAs limit the overall scale (e.g., 8-core processor).
Other emulation platforms connect many FPGAs into a single logical FPGA~\cite{DBLP:journals/tcad/BabbTDHHA97, fpga_emulation} to circumvent resource limits.
These systems share problems, such as partitioning, with software-based RTL simulation but also suffer from protracted compilation time (hours to days) to map logic to FPGA primitives.
However, they can emulate large systems as fast as 1~MHz at a high price.

\section{Conclusion}\label{sec:conclusion}

Thousand-way parallel RTL simulation is becoming necessary.
Current simulation techniques can adequately exploit only tens of cores in general-purpose processors because of their high synchronization and communication costs.

We used a 1472-core computer, the Graphcore IPU, to study the feasibility and challenges of massively parallel simulation.
Our study analyzed three dimensions of parallel simulation: synchronization, communication, and computation.
Using these results, we implemented \SimName, an RTL compiler that can use up to 5888 cores effectively.
Despite the IPU's almost $84\times$ single-core performance disadvantage against x64 machines, \SimName runs up to $4\times$ faster on large designs.

Our work demonstrates that thousand-way parallel RTL simulation is practical and beneficial.
It opens new avenues for future research that speeds up RTL simulation on massively parallel systems.
\section{Acknowledgements}
We are grateful to Graphcore for lending us the M2000 hardware.
We thank the the Graphcore staff and engineers, especially Mark Pupilli, Dario Domizioli, Peter Birch, Svetlomir Hristozkov, Marie-Ann Le Menn, and David Bozier.
They helped us throughout the project development, from getting us started to detailed explanations of the inner workings of \textcode{poplar} and \textcode{popc}, and providing feedback on our work.

At EPFL, Sahand Kashani and Rishabh Iyer's feedback on the writing and paper's narrative helped us make significant improvements.
Furthermore, Sanidhya Kashyap generously allowed us to use their machines to benchmark Verilator.
We thank Edouard Bugnion and Margaret Church for their support in the last year of the Very Large Scale Laboratory at EPFL.
Last but not least, we thank Jiacheng Ma, who made us realize the potential of using the IPUs for RTL simulation.



\bibliographystyle{plain}
\bibliography{./gen-abbrev,./dblp,./misc}

\begin{thebibliography}{10}

\bibitem{zen4}
4th gen {AMD} {EPYC} {P}rocessor {A}rchiecture.
\newblock Technical report, AMD.

\bibitem{gcore_ipu}
{AI} {IPU} {C}loud {I}nfrastructure.
\newblock \url{https://gcore.com/cloud/ai-platform}.
\newblock Accessed: 22-11-2023.

\bibitem{azure_instance_prices}
Azure pricing calculator.
\newblock \url{https://azure.microsoft.com/en-us/pricing/calculator/}.
\newblock Accessed 24-06-2024.

\bibitem{graphcore}
Introducing the {Colussus} {MK2 GC200} {IPU}.
\newblock \url{https://www.graphcore.ai/products/ipu}.
\newblock Accessed: 2023-11-23.

\bibitem{bitcoin}
{Open-Source FPGA Bitcoin Miner}.
\newblock \url{https://github.com/progranism/Open-Source-FPGA-Bitcoin-Miner}.

\bibitem{poplar_sdk}
{Poplar Graph Programming Framework}.
\newblock
  \url{https://docs.graphcore.ai/en/latest/child-pages/poplar.html#poplar}.

\bibitem{DBLP:conf/isca/AbtsKLKBBPADHTB22}
Dennis Abts, Garrin Kimmell, Andrew~C. Ling, John Kim, Matthew Boyd, Andrew
  Bitar, Sahil Parmar, Ibrahim Ahmed, Roberto DiCecco, David Han, John
  Thompson, Michael Bye, Jennifer Hwang, Jeremy Fowers, Peter Lillian, Ashwin
  Murthy, Elyas Mehtabuddin, Chetan Tekur, Thomas Sohmers, Kris Kang, Stephen
  Maresh, and Jonathan Ross.
\newblock {A software-defined tensor streaming multiprocessor for large-scale
  machine learning.}
\newblock In {\em Proceedings of the 49th International Symposium on Computer
  Architecture (ISCA)}, pages 567--580, 2022.

\bibitem{DBLP:conf/isca/AbtsRSWBHBTKKHL20}
Dennis Abts, Jonathan Ross, Jonathan Sparling, Mark Wong-VanHaren, Max Baker,
  Tom Hawkins, Andrew Bell, John Thompson, Temesghen Kahsai, Garrin Kimmell,
  Jennifer Hwang, Rebekah Leslie-Hurd, Michael Bye, E.~R. Creswick, Matthew
  Boyd, Mahitha Venigalla, Evan Laforge, Jon Purdy, Purushotham Kamath, Dinesh
  Maheshwari, Michael Beidler, Geert Rosseel, Omar Ahmad, Gleb Gagarin, Richard
  Czekalski, Ashay Rane, Sahil Parmar, Jeff Werner, Jim Sproch, Adrian Macias,
  and Brian Kurtz.
\newblock {Think Fast: A Tensor Streaming Processor (TSP) for Accelerating Deep
  Learning Workloads.}
\newblock In {\em Proceedings of the 47th International Symposium on Computer
  Architecture (ISCA)}, pages 145--158, 2020.

\bibitem{ahmad_numa}
Tariq~Bashir Ahmad, Namdo Kim, Byeong Min, Apurva Kalia, Maciej Ciesielski, and
  Seiyang Yang.
\newblock Scalable parallel event-driven {HDL} simulation for multi-cores.
\newblock In {\em 2012 International Conference on Synthesis, Modeling,
  Analysis and Simulation Methods and Applications to Circuit Design (SMACD)},
  pages 217--220, 2012.

\bibitem{DBLP:journals/micro/AmidBGGKLMMOPRS20}
Alon Amid, David Biancolin, Abraham Gonzalez, Daniel Grubb, Sagar Karandikar,
  Harrison Liew, Albert Magyar, Howard Mao, Albert~J. Ou, Nathan Pemberton,
  Paul Rigge, Colin Schmidt, John~Charles Wright, Jerry Zhao, Yakun~Sophia
  Shao, Krste Asanovic, and Borivoje Nikolic.
\newblock {Chipyard: Integrated Design, Simulation, and Implementation
  Framework for Custom SoCs.}
\newblock {\em IEEE Micro}, 40(4):10--21, 2020.

\bibitem{rocket_chip}
Krste Asanovi\'{c}, Rimas Avi\v{z}ienis, Jonathan Bachrach, Scott Beamer, David
  Biancolin, Christopher Celio, Henry Cook, Palmer Dabbelt, John Hauser, Adam
  Izraelevitz, Sagar Karandikar, Benjamin Keller, Donggyu Kim, John Koenig,
  Yunsup Lee, Eric Love, Martin Maas, Albert Magyar, Howard Mao, Miquel Moreto,
  Albert Ou, David Patterson, Brian Richards, Colin Schmidt, Stephen Twigg, Huy
  Vo, and Andrew Waterman.
\newblock The {Rocket} {Chip} {Generator}.
\newblock Technical report, University of California, Berkeley, 2016.

\bibitem{DBLP:journals/tcad/BabbTDHHA97}
Jonathan Babb, Russell Tessier, Matthew Dahl, Silvina Hanono, David~M. Hoki,
  and Anant Agarwal.
\newblock {Logic emulation with virtual wires.}
\newblock {\em IEEE Trans. Comput. Aided Des. Integr. Circuits Syst.},
  16(6):609--626, 1997.

\bibitem{DBLP:journals/micro/Beamer20}
Scott Beamer.
\newblock {A Case for Accelerating Software {RTL} Simulation}.
\newblock {\em {IEEE} Micro}, 40(4):112--119, 2020.

\bibitem{DBLP:conf/dac/BeamerD20}
Scott Beamer and David Donofrio.
\newblock Efficiently exploiting low activity factors to accelerate {RTL}
  simulation.
\newblock In {\em 57th {ACM/IEEE} Design Automation Conference, {DAC} 2020, San
  Francisco, CA, USA, July 20-24, 2020}, pages 1--6. {IEEE}, 2020.

\bibitem{nexus_emulation:22}
Peter Birch.
\newblock {Open source {FPGA}-based emulation with Nexus}.
\newblock In {\em Workshop on Open-Source EDA Technology (WOSET)}, number~1,
  2022.

\bibitem{DBLP:conf/dac/ChatterjeeDB09}
Debapriya Chatterjee, Andrew DeOrio, and Valeria Bertacco.
\newblock {Event-driven gate-level simulation with GP-GPUs}.
\newblock In {\em Proceedings of the 46th Design Automation Conference, {DAC}
  2009, San Francisco, CA, USA, July 26-31, 2009}, pages 557--562. {ACM}, 2009.

\bibitem{DBLP:conf/date/ChatterjeeDB09}
Debapriya Chatterjee, Andrew DeOrio, and Valeria Bertacco.
\newblock {{GCS:} High-performance gate-level simulation with GPGPUs}.
\newblock In Luca Benini, Giovanni~De Micheli, Bashir~M. Al{-}Hashimi, and
  Wolfgang M{\"{u}}ller, editors, {\em Design, Automation and Test in Europe,
  {DATE} 2009, Nice, France, April 20-24, 2009}, pages 1332--1337. {IEEE},
  2009.

\bibitem{DBLP:journals/todaes/ChatterjeeDB11}
Debapriya Chatterjee, Andrew DeOrio, and Valeria Bertacco.
\newblock {Gate-Level Simulation with GPU Computing.}
\newblock {\em ACM Trans. Design Autom. Electr. Syst.}, 16(3):30:1--30:26,
  2011.

\bibitem{DBLP:journals/sigarch/ChenAD09}
Jianwei Chen, Murali Annavaram, and Michel Dubois.
\newblock {SlackSim: a platform for parallel simulations of CMPs on CMPs.}
\newblock {\em SIGARCH Comput. Archit. News}, 37(2):20--29, 2009.

\bibitem{ash}
Fares Elsabbagh, Shabnam Sheikhha, Victor~A. Ying, Quan~M. Nguyen, Joel~S.
  Emer, and Daniel S{\'{a}}nchez.
\newblock {Accelerating {RTL} Simulation with Hardware-Software Co-Design}.
\newblock In {\em Proceedings of the 56th Annual {IEEE/ACM} International
  Symposium on Microarchitecture, {MICRO} 2023, Toronto, ON, Canada, 28 October
  2023 - 1 November 2023}, pages 153--166. {ACM}, 2023.

\bibitem{manticore_arxiv}
Mahyar Emami, Sahand Kashani, Keisuke Kamahori, Mohammad~Sepehr Pourghannad,
  Ritik Raj, and James~R. Larus.
\newblock {Manticore: Hardware-Accelerated RTL Simulation with Static
  Bulk-Synchronous Parallelism}.
\newblock In {\em Proceedings of the 28th ACM International Conference on
  Architectural Support for Programming Languages and Operating Systems, Volume
  4}, ASPLOS '23, page 219–237, New York, NY, USA, 2024. Association for
  Computing Machinery.

\bibitem{SiemensVerifStudyPart4Fpga}
Harry Foster.
\newblock {Part 4: The 2020 Wilson Research Group Functional Verification
  Study, FPGA Verification Effort Trends}, 12 2020.

\bibitem{SiemensVerifStudyPart8Asic}
Harry Foster.
\newblock {Part 8: The 2020 Wilson Research Group Functional Verification
  Study, IC/ASIC Resource Trends}, 1 2021.

\bibitem{DBLP:conf/ispass/FuW14}
Yaosheng Fu and David Wentzlaff.
\newblock {PriME: A parallel and distributed simulator for thousand-core
  chips.}
\newblock In {\em Proceedings of the 2014 IEEE International Symposium on
  Performance Analysis of Systems and Software (ISPASS)}, pages 116--125, 2014.

\bibitem{DBLP:journals/ior/GareyGJ78}
M.~R. Garey, Ronald~L. Graham, and David~S. Johnson.
\newblock {Performance Guarantees for Scheduling Algorithms.}
\newblock {\em Oper. Res.}, 26(1):3--21, 1978.

\bibitem{Herbst2024SwitchboardAO}
Steven Herbst, Noah Moroze, Edgar Iglesias, and Andreas Olofsson.
\newblock {Switchboard: An Open-Source Framework for Modular Simulation of
  Large Hardware Systems}.
\newblock {\em ArXiv}, abs/2407.20537, 2024.

\bibitem{DBLP:conf/iccad/IzraelevitzKLLW17}
Adam~M. Izraelevitz, Jack Koenig, Patrick Li, Richard Lin, Angie Wang, Albert
  Magyar, Donggyu Kim, Colin Schmidt, Chick Markley, Jim Lawson, and Jonathan
  Bachrach.
\newblock {Reusability is {FIRRTL} ground: Hardware construction languages,
  compiler frameworks, and transformations}.
\newblock In {\em 2017 {IEEE/ACM} International Conference on Computer-Aided
  Design, {ICCAD} 2017, Irvine, CA, USA, November 13-16, 2017}, pages 209--216.
  {IEEE}, 2017.

\bibitem{DBLP:conf/micro/JeffreySYES15}
Mark~C. Jeffrey, Suvinay Subramanian, Cong Yan, Joel~S. Emer, and Daniel
  Sánchez.
\newblock {A scalable architecture for ordered parallelism.}
\newblock In {\em Proceedings of the 48th Annual IEEE/ACM International
  Symposium on Microarchitecture (MICRO)}, pages 228--241, 2015.

\bibitem{DBLP:journals/corr/abs-1912-03413}
Zhe Jia, Blake Tillman, Marco Maggioni, and Daniele~Paolo Scarpazza.
\newblock {Dissecting the Graphcore IPU Architecture via Microbenchmarking.}
\newblock {\em CoRR}, abs/1912.03413, 2019.

\bibitem{DBLP:journals/micro/KarandikarMKBAL19}
Sagar Karandikar, Howard Mao, Donggyu Kim, David Biancolin, Alon Amid, Dayeol
  Lee, Nathan Pemberton, Emmanuel Amaro, Colin Schmidt, Aditya Chopra, Qijing
  Huang, Kyle Kovacs, Borivoje Nikolic, Randy~Howard Katz, Jonathan Bachrach,
  and Krste Asanovic.
\newblock {FireSim: FPGA-Accelerated Cycle-Exact Scale-Out System Simulation in
  the Public Cloud.}
\newblock {\em IEEE Micro}, 39(3):56--65, 2019.

\bibitem{DBLP:conf/date/KimCY11}
Dusung Kim, Maciej~J. Ciesielski, and Seiyang Yang.
\newblock A new distributed event-driven gate-level {HDL} simulation by
  accurate prediction.
\newblock In {\em Design, Automation and Test in Europe, {DATE} 2011, Grenoble,
  France, March 14-18, 2011}, pages 547--550. {IEEE}, 2011.

\bibitem{fpga_emulation}
Helena Krupnova and Gabriele Saucier.
\newblock {FPGA}-based emulation: Industrial and custom prototyping solutions.
\newblock In {\em Proceedings of the The Roadmap to Reconfigurable Computing,
  10th International Workshop on Field-Programmable Logic and Applications},
  FPL '00, page 68–77, Berlin, Heidelberg, 2000. Springer-Verlag.

\bibitem{DBLP:journals/micro/Lauterbach21}
Gary Lauterbach.
\newblock {The Path to Successful Wafer-Scale Integration: The Cerebras Story}.
\newblock {\em {IEEE} Micro}, 41(6):52--57, 2021.

\bibitem{DBLP:conf/vlsid/LiGL04}
Tun Li, Yang Guo, and Sikun Li.
\newblock {Design and Implementation of a Parallel Verilog Simulator: PVSim.}
\newblock In {\em VLSI Design}, pages 329--334, 2004.

\bibitem{rtlflow}
Dian{-}Lun Lin, Haoxing Ren, Yanqing Zhang, Brucek Khailany, and Tsung{-}Wei
  Huang.
\newblock {From {RTL} to {CUDA:} {A} {GPU} Acceleration Flow for {RTL}
  Simulation with Batch Stimulus}.
\newblock In {\em Proceedings of the 51st International Conference on Parallel
  Processing, {ICPP} 2022, Bordeaux, France, 29 August 2022 - 1 September
  2022}, pages 88:1--88:12. {ACM}, 2022.

\bibitem{DBLP:conf/date/LopezParadisLAWMB23}
Guillem L{\'{o}}pez{-}Parad{\'{\i}}s, Brian Li, Adri{\`{a}} Armejach, Stefan
  Wallentowitz, Miquel Moret{\'{o}}, and Jonathan Balkind.
\newblock {Fast Behavioural {RTL} Simulation of 10B Transistor {SoC} Designs
  with Metro-Mpi}.
\newblock In {\em Design, Automation {\&} Test in Europe Conference {\&}
  Exhibition, {DATE} 2023, Antwerp, Belgium, April 17-19, 2023}, pages 1--6.
  {IEEE}, 2023.

\bibitem{xorshift}
George Marsaglia.
\newblock Xorshift {RNG}s.
\newblock {\em Journal of Statistical Software}, 8(14):1–6, 2003.

\bibitem{DBLP:conf/hpca/MillerKKGBCEA10}
Jason~E. Miller, Harshad Kasture, George Kurian, Charles~Gruenwald III, Nathan
  Beckmann, Christopher Celio, Jonathan Eastep, and Anant Agarwal.
\newblock {Graphite: A distributed parallel simulator for multicores.}
\newblock In {\em Proceedings of the 16th IEEE Symposium on High-Performance
  Computer Architecture (HPCA)}, pages 1--12, 2010.

\bibitem{DBLP:journals/micro/MoreauCVRYZFJCG19}
Thierry Moreau, Tianqi Chen, Luis Vega, Jared Roesch, Eddie~Q. Yan, Lianmin
  Zheng, Josh Fromm, Ziheng Jiang, Luis Ceze, Carlos Guestrin, and Arvind
  Krishnamurthy.
\newblock {A Hardware-Software Blueprint for Flexible Deep Learning
  Specialization.}
\newblock {\em IEEE Micro}, 39(5):8--16, 2019.

\bibitem{DBLP:conf/aspdac/NanjundappaPJS10}
Mahesh Nanjundappa, Hiren~D. Patel, Bijoy~Antony Jose, and Sandeep~K. Shukla.
\newblock {SCGPSim: a fast SystemC simulator on GPUs}.
\newblock In {\em Proceedings of the 15th Asia South Pacific Design Automation
  Conference, {ASP-DAC} 2010, Taipei, Taiwan, January 18-21, 2010}, pages
  149--154. {IEEE}, 2010.

\bibitem{systemc}
OSCI.
\newblock {SystemC}.
\newblock \url{https://www.systemc.org}.

\bibitem{pico}
{PicoRV32 - A Size-Optimized RISC-V CPU}.
\newblock \url{https://github.com/YosysHQ/picorv32}.

\bibitem{DBLP:conf/iccad/QianD11}
Hao Qian and Yangdong Deng.
\newblock {Accelerating {RTL} simulation with GPUs}.
\newblock In Joel~R. Phillips, Alan~J. Hu, and Helmut Graeb, editors, {\em 2011
  {IEEE/ACM} International Conference on Computer-Aided Design, {ICCAD} 2011,
  San Jose, California, USA, November 7-10, 2011}, pages 687--693. {IEEE}
  Computer Society, 2011.

\bibitem{processor_scaling_data}
Karl Rupp.
\newblock Microprocessor trend data.
\newblock \url{https://github.com/karlrupp/microprocessor-trend-data}, 2022.
\newblock Accessed: 18-10-2023.

\bibitem{DBLP:journals/jacm/Sahni76}
Sartaj Sahni.
\newblock {Algorithms for Scheduling Independent Tasks.}
\newblock {\em J. ACM}, 23(1):116--127, 1976.

\bibitem{Sarkar86}
Vivek Sarkar and John~L. Hennessy.
\newblock {Compile-time partitioning and scheduling of parallel programs.}
\newblock In {\em SIGPLAN Symposium on Compiler Construction}, pages 17--26,
  1986.

\bibitem{DBLP:journals/jea/SchlagHGASS22}
Sebastian Schlag, Tobias Heuer, Lars Gottesbüren, Yaroslav Akhremtsev,
  Christian Schulz, and Peter Sanders.
\newblock {High-Quality Hypergraph Partitioning.}
\newblock {\em ACM J. Exp. Algorithmics}, 27:1.9:1--1.9:39, 2022.

\bibitem{snyder_verilator_2020}
Wilson Snyder.
\newblock Verilator, accelerated: Accelerating development, and case study of
  accelerating performance.
\newblock 2nd Workshop on Open-Source Design Automation ({OSDA}).

\bibitem{snyder_mt_2018}
Wilson Snyder.
\newblock Verilator 4.0: Open simulation goes multithreaded.
\newblock The OPen Source Digital Design Conference ({ORConf}), 2018.

\bibitem{snyder_mt_2019}
Wilson Snyder.
\newblock Your {Big} 4th {Simulator}: 2019 intro and roadmap.
\newblock {CHIPS Alliance}, 2019.

\bibitem{SLB}
Zoya Svitkina and Lisa Fleischer.
\newblock {Submodular Approximation: Sampling-based Algorithms and Lower
  Bounds.}
\newblock {\em SIAM J. Comput.}, 40(6):1715--1737, 2011.

\bibitem{DBLP:conf/isca/SanchezK13}
Daniel Sánchez and Christos Kozyrakis.
\newblock {ZSim: fast and accurate microarchitectural simulation of
  thousand-core systems.}
\newblock In {\em Proceedings of the 40th International Symposium on Computer
  Architecture (ISCA)}, pages 475--486, 2013.

\bibitem{DBLP:conf/asplos/TanQCAP15}
Zhangxi Tan, Zhenghao Qian, Xi~Chen, Krste Asanovic, and David~A. Patterson.
\newblock {DIABLO: A Warehouse-Scale Computer Network Simulator using FPGAs.}
\newblock In {\em Proceedings of the 20th International Conference on
  Architectural Support for Programming Languages and Operating Systems
  (ASPLOS-XX)}, pages 207--221, 2015.

\bibitem{DBLP:conf/fpt/TianB08}
Xiang Tian and Khaled Benkrid.
\newblock {Design and implementation of a high performance financial
  Monte-Carlo simulation engine on an {FPGA} supercomputer}.
\newblock In Tarek~A. El{-}Ghazawi, Yao{-}Wen Chang, Juinn{-}Dar Huang, and
  Proshanta Saha, editors, {\em 2008 International Conference on
  Field-Programmable Technology, {FPT} 2008, Taipei, Taiwan, December 7-10,
  2008}, pages 81--88. {IEEE}, 2008.

\bibitem{mpsched_np}
Jeffrey~D. Ullman.
\newblock {NP-Complete Scheduling Problems.}
\newblock {\em J. Comput. Syst. Sci.}, 10(3):384--393, 1975.

\bibitem{valiant_bridging_1990}
Leslie~G. Valiant.
\newblock {A Bridging Model for Parallel Computation.}
\newblock {\em Commun. ACM}, 33(8):103--111, 1990.

\bibitem{DBLP:conf/dac/VincoCBF12}
Sara Vinco, Debapriya Chatterjee, Valeria Bertacco, and Franco Fummi.
\newblock {SAGA:} systemc acceleration on {GPU} architectures.
\newblock In Patrick Groeneveld, Donatella Sciuto, and Soha Hassoun, editors,
  {\em The 49th Annual Design Automation Conference 2012, {DAC} '12, San
  Francisco, CA, USA, June 3-7, 2012}, pages 115--120. {ACM}, 2012.

\bibitem{repcut}
Haoyuan Wang and Scott Beamer.
\newblock {RepCut: Superlinear Parallel RTL Simulation with Replication-Aided
  Partitioning}.
\newblock In {\em Proceedings of the 28th ACM International Conference on
  Architectural Support for Programming Languages and Operating Systems, Volume
  3}, ASPLOS 2023, page 572–585, New York, NY, USA, 2023. Association for
  Computing Machinery.

\bibitem{DBLP:conf/dac/WangHPZ87}
L.{-}T. Wang, Nathan~E. Hoover, Edwin~H. Porter, and John~J. Zasio.
\newblock {SSIM:} {A} software levelized compiled-code simulator.
\newblock In A.~O'Neill and D.~Thomas, editors, {\em Proceedings of the 24th
  {ACM/IEEE} Design Automation Conference. Miami Beach, FL, USA, June 28 - July
  1, 1987}, pages 2--8. {IEEE} Computer Society Press / {ACM}, 1987.

\bibitem{DBLP:conf/dac/WangM90}
Zhicheng Wang and Peter~M. Maurer.
\newblock {LECSIM:} {A} levelized event driven compiled logic simulation.
\newblock In Richard~C. Smith, editor, {\em Proceedings of the 27th {ACM/IEEE}
  Design Automation Conference. Orlando, Florida, USA, June 24-28, 1990}, pages
  491--496. {IEEE} Computer Society Press, 1990.

\bibitem{DBLP:conf/date/YangHKKCPK14}
Seiyang Yang, Jaehoon Han, Doowhan Kwak, Namdo Kim, Daeseo Cha, Junhyuck Park,
  and Jay Kim.
\newblock {Predictive parallel event-driven {HDL} simulation with a new
  powerful prediction strategy}.
\newblock In Gerhard~P. Fettweis and Wolfgang Nebel, editors, {\em Design,
  Automation {\&} Test in Europe Conference {\&} Exhibition, {DATE} 2014,
  Dresden, Germany, March 24-28, 2014}, pages 1--3. European Design and
  Automation Association, 2014.

\bibitem{constellation}
Jerry Zhao, Animesh Agrawal, Borivoje Nikolic, and Krste Asanović.
\newblock Constellation: An open-source {SoC}-capable {NoC} generator.
\newblock In {\em 2022 15th IEEE/ACM International Workshop on Network on Chip
  Architectures (NoCArc)}, pages 1--7, 2022.

\bibitem{zhaosonicboom}
Jerry Zhao, Ben Korpan, Abraham Gonzalez, and Krste Asanovic.
\newblock {SonicBOOM: The 3rd Generation Berkeley Out-of-Order Machine}.
\newblock May 2020.

\bibitem{DBLP:conf/micro/Zhou0LWH23}
Kexing Zhou, Yun Liang, Yibo Lin, Runsheng Wang, and Ru~Huang.
\newblock {Khronos: Fusing Memory Access for Improved Hardware {RTL}
  Simulation}.
\newblock In {\em Proceedings of the 56th Annual {IEEE/ACM} International
  Symposium on Microarchitecture, {MICRO} 2023, Toronto, ON, Canada, 28 October
  2023 - 1 November 2023}, pages 180--193. {ACM}, 2023.

\bibitem{DBLP:conf/date/ZurstrassenCJYXL23}
Niko Zurstra{\ss}en, Jos{\'{e}} Cubero{-}Cascante, Jan~Moritz Joseph,
  Li~Yichao, Xinghua Xie, and Rainer Leupers.
\newblock {par-gem5: Parallelizing gem5's Atomic Mode}.
\newblock In {\em Design, Automation {\&} Test in Europe Conference {\&}
  Exhibition, {DATE} 2023, Antwerp, Belgium, April 17-19, 2023}, pages 1--6.
  {IEEE}, 2023.

\end{thebibliography}

\end{document}